# Partial covariance two-dimensional mass spectrometry for determination of biomolecular primary structure

Taran Driver[1], Ruth Ayers[1], Rüdiger Pipkorn[3], Bridgette Cooper[1], Nikhil Bachhawat[1], Serguei Patchkovskii[4], Vitali Averbukh[1], David R. Klug[2], Jon P. Marangos[1], Leszek J. Frasinski[1], and Marina Edelson-Averbukh[1,2,*]

[1]*The Blackett Laboratory Laser Consortium, Department of Physics, Imperial College London, London, SW7 2AZ, United Kingdom*
[2]*Department of Chemistry Imperial College London, London, SW7 2AZ, United Kingdom*
[3]*German Cancer Research Centre, Department of Translational Immunology, INF 580, 69120 Heidelberg, Germany*
[4]*Max Born Institute for Nonlinear Optics and Short Pulse Spectroscopy, Max-Born-Straße 2A, 12489 Berlin, Germany*
*e-mail: m.edelson-averbukh@imperial.ac.uk

**Mass spectrometry (MS) is used widely in biomolecular structural analysis and is particularly dominant in the study of proteins. Despite its considerable power, state-of-the-art protein MS frequently suffers from limited reliability of spectrum-to-structure assignments. This could not be solved fully by the dramatic increase in mass accuracy and resolution of modern MS instrumentation or by the introduction of new fragmentation methods. Here we present a new kind of two-dimensional mass spectrometry for high fidelity determination of a biomolecular primary structure based on partial covariance mapping. Partial covariance two-dimensional mass spectrometry (pC-2DMS) detects intrinsic statistical correlations between biomolecular fragments originating from the same or consecutive decomposition events. This enables identification of pairs of ions produced along the same fragmentation pathway of a biomolecule across its entire fragment mass spectrum. We demonstrate that the *fragment-fragment* correlations revealed by pC-2DMS provide much more specific information on the amino acid sequence and its covalent modifications than the individual fragment mass-to-charge ratios on which standard one-dimensional MS is based. We illustrate the power of pC-2DMS by using it to resolve structural isomers of combinatorially modified histone peptides inaccessible to standard MS.**

Mass spectrometry is the method of choice for the structural analysis of biomolecules, such as proteins[1,2], nucleic acids, lipids and metabolites. The primary aim of protein MS analysis is to identify a protein by establishing the sequence of its molecular building blocks, i.e. amino acids, and to characterise possible post-translational modifications (PTMs)[3]. To do so, the biomolecules are typically first cut into smaller fragments, e.g. using enzymes to obtain peptides, which are subsequently sent to a mass spectrometer via soft ionisation techniques [electrospray ionisation (ESI)[4] or matrix-assisted laser desorption-ionisation (MALDI)[5]]. There, they are usually further fragmented to obtain detailed structural information by activation methods of tandem mass spectrometry (MS/MS), the most prevalent being collision-induced dissociation (CID)[6], and finally the spectral information is pieced together to deduce the primary structure of the original biomolecule.

The crucial step of the protein MS workflow is the deduction of the amino acid sequence and its PTMs from the tandem mass spectra. This task can be accomplished using a range of approaches which rely on either matching the experimental spectra to "theoretical" ones[7] derived from protein databases and a set of generalised peptide fragmentation rules[8,9], matching the acquired MS/MS spectra to spectral libraries[10], or performing a first-principles structural reconstruction using the measured spectrum and fragmentation rules only (so-called *de novo* algorithms)[11]. Whichever method is chosen for the data interpretation, normally less than 60% of the measured fragment mass spectra are successfully interpreted and matched to the correct peptide and protein sequences[12–14]. Failure in spectrum-to-structure assignment is caused by a number of factors, among them the strong variability of peptide decomposition patterns as a function of the analysed amino acid sequence, presence of PTMs, type and location of the modifying groups, peptide length, charge state etc.[15–17], frequently leading to considerable deviation of the experimental mass spectra from those predicted by the simplified peptide fragmentation rules. Moreover, the MS analysis is compromised by false identifications caused by fragments attributed to isobaric (within the given mass tolerance), isomeric or even identical ions belonging to incorrect structures, the most prominent example of the latter being structural characterisation of co-fragmented combinatorially modified peptides[18,19].

Mainstream one-dimensional (1D) MS is striving to overcome the spectral assignment challenges through two main avenues: (i) the development of new fragmentation methods[20–22] and (ii) improving the accuracy and resolution of the mass-to-charge (*m/z*) measurement[14,23]. For example, the introduction of electron-based activation techniques, such as electron transfer (ETD)[20] and electron capture dissociation (ECD)[21], which randomly cleave the peptide backbone bonds, has led to a crucial increase in sequence coverage compared to the traditional CID technique, thus providing a higher structural specificity of the obtained fragment mass spectra. Furthermore, the combined use of electron- and collision-based dissociation methods has achieved an additional enhancement in the reliability of spectrum-to-structure assignments[24,25]. The remarkable resolving power and mass accuracy of ion cyclotron resonance (ICR) and the Orbitrap mass analysers allowed one to measure ions at sub part-per-million mass accuracy, boosting considerably the structural specificity of the *m/z* measurement[26].

However, the most advanced forms of 1D mass spectrometry, that unevitably rely on individual fragment mass-to-charge ratios, cannot solve some central problems of biomolecular MS, even at complete sequence coverage and theoretically infinite mass accuracy. An example, as we show here, is the formidable analytical challenge of identifying co-fragmented isomeric combinatorially modified peptides, which often lack any therotetically possible isomer-specific fragments by which the presence of particular isomers could have been detected[18,19,27]. Within pC-2DMS, we solve this topical problem by using statistical correlations between fragment ion peaks across a series of repeated MS/MS measurements (see Fig. 1) to establish isomer-specific structural correlations that cannot be obtained by any existing MS approach. Noteworthy, the only currently available form of two-dimensional MS, 2D Fourier transform (FT) ICR[28,29] is inherently unable to solve such structural problems, because of the lack of the *fragment*-fragment correlation information. Indeed, in contrast to pC-2DMS, the 2D FT-ICR technique is based on the same physical principle as 2D exchange NMR spectroscopy and produces *precursor*-fragment ion correlations, that can otherwise be generated through 1D MS/MS approach provided high enough precursor isolation efficiency[6].

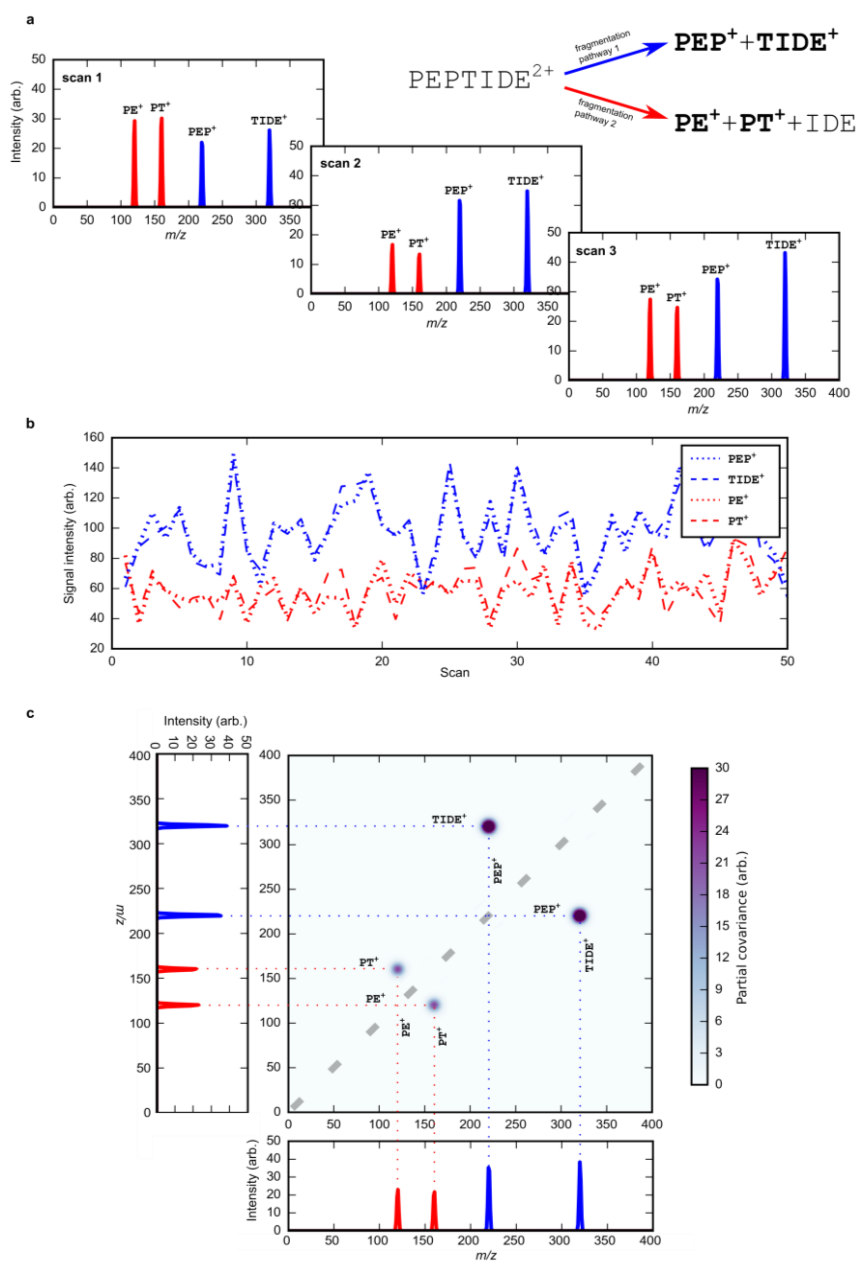

**Figure 1 | The principle of pC-2DMS.** The inherent scan-to-scan fluctuations in the abundances of fragments born in the same (panel a) and consecutive (e.g. "niblings" formation, see text) reactions follow each other in a series of individual MS/MS scans (panel b). pC-2DMS exploits this property to identify such fragment pairs, by analysing a series of repeated MS/MS scans and calculating the partial covariance between all fragment intensities. The correlating fragment pairs are joined by positive islands on the pC-2DMS map (panel c). The map is symmetric with respect to the $x = y$ autocorrelation diagonal (grey dashed line), because pC-2DMS correlates the

same fragment mass spectrum with itself. The autocorrelation line contains uninformative self-correlations of each spectral signal and is usually removed for visual clarity, as it has been done here.

pC-2DMS is based on the physical principle of partial covariance mapping[30]. Covariance mapping spectroscopy was developed by Frasinski *et al*. as a tool for the study of mechanisms of radiation-induced fragmentation of di- and tri-atomic molecules[31]. The technique is based on measuring the covariance, $Cov(X,Y)$, between the intensities of every pair of spectral signals *X* and *Y* across a series of ionic time-of-flight spectra,

$$Cov(X,Y) = \langle XY \rangle - \langle X \rangle \langle Y \rangle, \qquad (1)$$

where angular brackets denote averaging over multiple spectra. If a pair of signals, *X* and *Y*, are characterised by positive covariance, their intensities fluctuate synchronously across different spectra, and as a consequence the corresponding fragments X and Y are determined to originate from the same decomposition process, e.g. Z→X+Y. Covariance mapping has been shown to be effective, for example, in unravelling the decomposition mechanisms of 'hollow atoms' – unstable states of matter formed by intense X-ray free electron laser (XFEL) irradiation – or in correlating photoelectron emission with the fragmentation of hydrocarbons in intense infrared laser fields[32]. An extension of covariance mapping, known as 'partial covariance mapping', was developed to single out the mechanistic correlations from extrinsic (uninteresting) ones, which affect each and every spectral signal simultaneously in a uniform manner and can result, for example, from fluctuations in laser intensity[32] (see also Supplementary Information). Partial covariance mapping has allowed researchers to study X-ray induced fragmentation at XFELs under significant intensity fluctuations of the ionising pulse[33]. All the covariance mapping studies have been so far performed to elucidate the laser-induced ionisation and/or decomposition mechanisms of atoms or small molecules [e.g. $N_2$ & $I_2$[33]] of *a priori* well known structure. In what follows, we use partial covariance mapping concept for the first time to deduce the unknown primary structure of a biomolecule, demonstrating the pC-2DMS technique and its advantages over conventional analytical MS.

## Results and Discussion

**Total ion current partial covariance mapping**

In a typical biomolecular MS measurement, there exist multiple fluctuating experimental parameters affecting all the fragment intensities simultaneously. For example, in the case of the ion trap ESI CID experiments performed in this study, these are ion focusing voltages, axial and radial trapping voltages, resonance excitation and ejection voltages, ion trap helium gas pressure, electrospray flow rate and others. Monitoring changes in all, or at least in a series of the most dominant (if identified) experimental parameters causing spectral fluctuations on a scan-to-scan basis, as required by the standard partial covariance formulation[32], is impractical and frequently not possible. Within pC-2DMS, this obstacle has been overcome through development of a new form of partial covariance mapping that is based on the single, readily accessible parameter, namely the total ion count (TIC) of a fragment mass spectrum. We call this approach total ion current partial covariance mapping. Indeed, since the compound effect of all fluctuations in the experimental conditions influences the total number of MS/MS scan fragment ions, the TIC partial covariance, $pCov(X,Y,TIC)$, enables us to extract true fragment ion correlations between fragments born in the same/consecutive molecular decompositions while suppressing the otherwise overwhelming extrinsic correlations dominating the simple covariance map (shown in Supplementary Fig. 1). The TIC partial covariance between each pair of fragment mass spectral signals *X* and *Y* is given by:

$$pCov(X,Y,TIC) = Cov(X,Y) - Cov(X,TIC)Cov(Y,TIC)/Cov(TIC,TIC) \qquad (2)$$

In order to provide a theoretical basis for our TIC partial covariance mapping procedure we have expanded the statistical theory of covariance for noisy Poissonian processes[34]. This theoretical framework is valid when the number of ions trapped in each scan follows a noise-augmented Poisson distribution, as we have verified for our typical ESI experiments. The extended theory gives the TIC partial covariance between two fragment ions X and Y for the fragmentation of biopolymers as:

$$pCov(X,Y,TIC) = \beta^2 \nu_0 \alpha_1 - \omega \nu_0, \qquad (3)$$

where $\nu_0$ is the average number of parent ions fragmented at each scan, $\alpha_1$ is the probability that a single parent ion will dissociate to produce fragments X and Y, and $\beta$ is the instrument fragmentation-to-detection efficiency, typically required to be ~30 % or higher for successful covariance analysis[32]. The small additional term $-\omega v_0$ arises due to the imperfection of the TIC-based partial-covariance correction, is guaranteed to be negative, see Supplementary Information. Eq. 3 shows that the TIC partial covariance between X and Y is linearly dependent on parent ion abundance and the branching ratio of the process that produces X and Y together. This linearity renders pC-2DMS suitable not only for qualitative, but also for quantitative analysis. Negative TIC partial covariance between fragments X and Y does not stem from them being born in the same or consecutive decomposition reactions and can be disregarded.

**pC-2DMS – demonstration of principle**

We illustrate how pC-2DMS works on a triply protonated acetylated peptide of human histone H3.1 protein [$^{118}$VTIMPK$_{Ac}$DIQLAR$^{129}$, K$_{Ac}$ = N$^{\varepsilon}$-acetyllysine], **P1**. A conventional 1D CID mass spectrum of **P1** displays a wide range of standard sequence-specific fragments, so-called a-, b- and y-ions[35], providing a good model for the pC-2DMS demonstration. The pC-2DMS map, which shows the value of the TIC partial covariance (Eq. 2) between every pair (*X, Y*) of CID spectral signals, is presented in Fig. 2. By construction, pC-2DMS maps are symmetric with respect to the main diagonal ($m/z_x=m/z_y$, autocorrelation diagonal) along which all fragment signals trivially correlate with themselves (we remove it for clarity). The off-diagonal peaks of Fig. 2 reveal correlations between pairs of peptide fragments (X, Y) of shared origin generated in the same or in consecutive decompositions of the ionised peptide. The geometric position of the off-diagonal correlation peaks on the pC-2DMS map defines three general types of signals. First is the correlation peaks of pairs of complementary fragments (plotted in blue, e.g. $y_8^{2+}$ & $b_4^+$) generated by fission of a single particular bond of the peptide, which arrange themselves along so-called mass conservation lines (plotted as dashed and dotted lines in Fig. 2). For the correlation signals lying along the dashed line, the fragment measured at $m/z_x$ has charge 1+ whilst its sibling measured at $m/z_y$ has charge 2+, and *vice-versa* for the dotted line (see e.g. the $y_6^+$ & $b_6^{2+}$ signal labelled on both mass-conservation lines by ✴). The charge state partition for these pC-2DMS signals can be directly calculated from the gradient of the mass conservation line. Second is the correlation signals of fragments originating from consecutive decompositions through loss of neutrals, showing as horizontal and vertical peak series relative to the pC-2DMS peaks of the corresponding intact ions (plotted in green, e.g. $y_9^{2+}$ & $[b_3-H_2O]^+$). Third is the manifold of scattered signals revealing correlations of consecutive charge loss products such as terminus-free internal fragments and their derivatives[36] (plotted in red, e.g. $b_{i(2-8)}^+$ & $y_4^+$ and $[b_{i(5-8)}-CO]^+$ & $y_4^+$) and non-complementary terminal ions (magenta, e.g. $y_2^+$ & $b_9^+$). The pairs of fragments revealed by pC-2DMS can be classified as "*siblings*" (Z→**X**+**Y**), "*niblings*" (Z→**X**+Y$_1$, Y$_1$→X$_1$+**Y**), "*cousins*" (Z→X$_1$+Y$_1$, X$_1$→**X**+X$_2$, Y$_1$→**Y**+Y$_2$) etc.

The fragment-fragment correlations of **P1** revealed by pC-2DMS (Fig. 2) provide direct experimental evidence for the formation history of the fragments, enabling reliable assignment of both terminal and internal mass spectral signals to the correct peptide sequence. Thus for example, the internal fragment $b_{i(5-8)}^+$ [PK$_{Ac}$DI]$^+$ (*m/z* 496.3) can be rapidly recognised based on the observed correlation of both itself and its CO-loss derivative with the corresponding N- and C- terminal parts of **P1**, $b_4^+$ [VTIM]$^+$ and $y_4^+$ [QLAR]$^+$ (see signals Ⓐ-Ⓒ), the identities of which are in turn confirmed through the correlations with the sibling and nibling ions (signals Ⓓ and Ⓔ). Another example is the peptide granddaughter fragment at *m/z* 369.2, generated by uncommon methionine-specific elimination of methythioethane from the [VTIM]$^+$ ($b_4^+$) ion, that can be confidently assigned based on its correlation with the nibling ion *m/z* 491.8 [PK(Ac)DIQLAR]$^{2+}$ ($y_8^{2+}$), see signal Ⓕ in Fig. 2. The sequence-specific fragment-fragment correlations of Fig. 2 experimentally demonstrate the validity of the TIC partial covariance introduced in Eq. 2. They also indicate that covariance mapping can be performed both for trapped ions and for species as large as peptides with masses of the order of kDa. Finally, the data of Fig. 2 show that the correlation of consecutive decomposition products, and not only of the sibling fragments, can be revealed by covariance mapping. We have successfully tested pC-2DMS on a variety of unmodified and PTM-containing peptide sequences, as well as on RNA and DNA oligonucleotides and whole proteins (see Supplementary Information Table 3 and Figs. 9-16 for representative examples). Our data indicate that the pC-2DMS principle is generally applicable for structural analysis of diverse biological polymers. Noteworthy, the pC-2DMS fragment ion correlations greatly facilitate the assignment of such anomalous biomolecular fragments as those produced by scrambling of the original primary structure (see Supplementary Fig. 8).

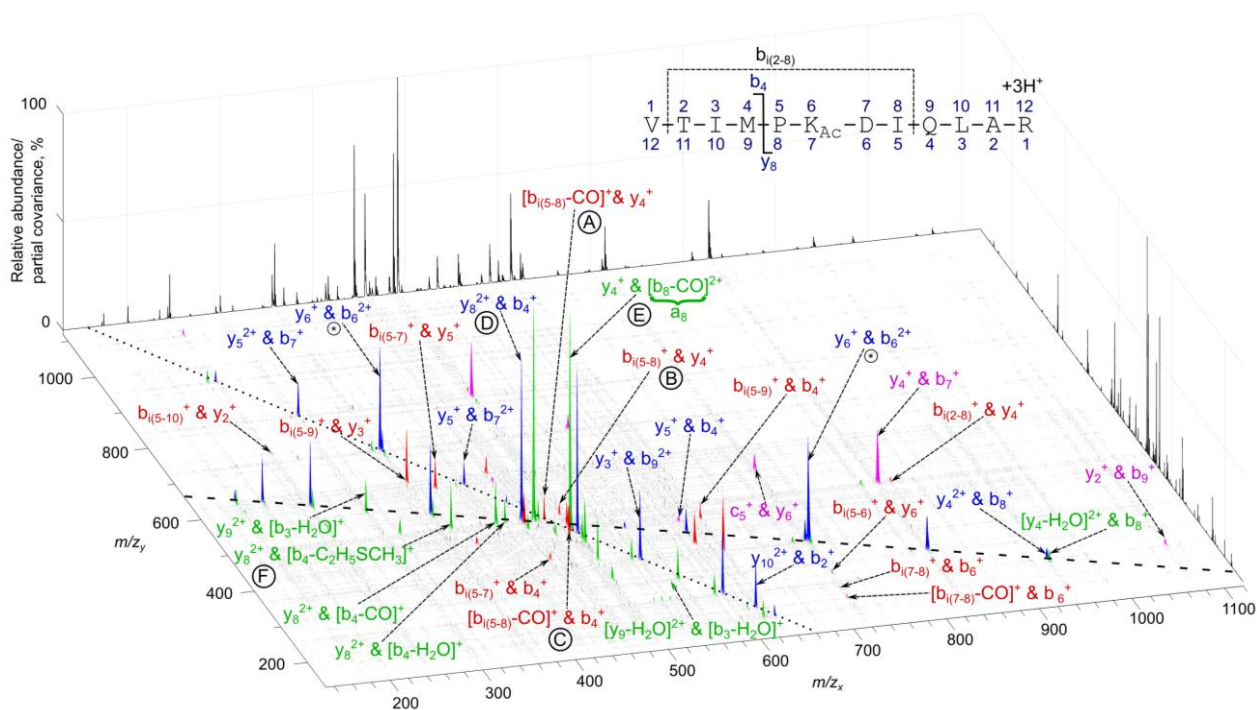

**Figure 2 | pC-2DMS map of triply protonated peptide VTIMPK$_{Ac}$DIQLAR (P1).** The vertical axis represents the relative value of $pCov(X,Y,TIC)$; the axes $m/z_x$ and $m/z_y$ display the correlating fragment mass-to-charge ratios. The autocorrelation diagonal has been removed for clarity. The black line graph plotted against the back walls of the map is the averaged 1D CID spectrum of **P1**. Selected pC-2DMS peaks are annotated (the peptide fragment nomenclature used[18] is given in the inset) and the signal colours represent the correlating fragment identities: blue – complementary ions, green – secondary neutral loss products, red – one or more terminus-free internal ions, magenta – one or more non-complementary terminal fragments. The signals marked by Ⓐ-Ⓕ are discussed in the text. Apart from the $y_6^+$ & $b_6^{2+}$ signal (marked by ⊛), the pC-2DMS peaks are annotated only on one side of the autocorrelation diagonal for clarity. The mass conservation lines $[m([\mathbf{P1}+3H]^{3+}) = 1\cdot m/z_x + 2\cdot m/z_y]$ and $[m([\mathbf{P1}+3H]^{3+}) = 2\cdot m/z_x + 1\cdot m/z_y]$, where $m([\mathbf{P1}+3H]^{3+})$ is the protonated peptide mass, are shown by dashed and dotted lines, respectively.

The fragment-fragment connectivity embodied in pC-2DMS signals render them much more specific to the biomolecular sequence than the corresponding 1D MS fragment ion peaks. For example, in the case of correlations involving internal ions of the length typical of tryptic peptides, we have shown that the level of structural specificity of pC-2DMS correlations obtained at a modest 0.8 Da *m/z* tolerance cannot be achieved by 1D MS even at infinite mass accuracy, for any fragment ion type (see Supplementary Fig. 6). pC-2DMS uncovers the structure-specific fragments spanning more than three orders of magnitude in relative abundance (see Supplementary Information, Fig. 3). In order to differentiate very low intensity pC-2DMS true signals from possible statistical noise, we introduce a correlation score *S(X,Y)* calculated by normalising the volume of pC-2DMS signals (*X*, *Y*) to the standard deviation $\sigma(V)$ of their volume upon jackknife resampling[37]:

$$S(X,Y) = \frac{V[pCov(X,Y;I)]}{\sigma(V)}, \qquad (4)$$

We rank all the pC-2DMS signals according to their relative correlation scores taken as a percentage of the highest *S(X,Y)* off-diagonal peak on the map. The *S(X,Y)* score reflects the stability of the pC-2DMS correlation signals and turns out to be superior to pC-2DMS peak height or volume (defined by relative value of TIC partial covariance, see Fig. 2) as a measure of true structural signals, see Supplementary Information Fig. 2. The introduced correlation score can be used as a universal parameter for pC-2DMS peak selection across the full *m/z* range in an automatic pC-2DMS spectrum-to-structure matching engine. We have designed and successfully tested (Supplementary Fig. 7) a prototype of an automated pC-2DMS database search engine, whose algorithm is based exclusively on fragment-fragment correlations selected and weighted by their correlation scores.

## pC-2DMS for combinatorial PTMs

We demonstrate the analytical power of pC-2DMS by using it to solve an outstanding problem of state-of-the-art MS, namely the analysis of combinatorial modification patterns, considering post-translational modifications on histone peptides as a particular example. The major challenge of such analysis [e.g. for studying the 'histone code'[38]] is resolving mixtures of co-fragmented positional isomers which feature the same modified residues but in different locations within the peptide sequence. This problem is caused by the frequent absence of unique ("marker") 1D MS fragments, irrespective of the fragmentation method, sensitivity, mass accuracy or resolution, which would enable one to distinguish each isomer from every other co-fragmented structure. Indeed, resolution of the series of diacetylated histone H4 peptide isomers which we resolve here was previously considered as "mathematically impossible" (see '3AC' isomers in Fig. 2 of reference 27; note that the fixed N-terminal acetylation does not contribute to the combinatorial modification patterns[27]). Effectiveness of the currently available methods for combinatorial isomer characterisation is strongly limited either by pre-analysis separation capability[18] or by the restrictive requirement of targeted comparison of the fragment MS data to the individually obtained spectra of every possible isomer in its pure form[19]. Here we concentrate on mixtures of the co-fragmented isomeric diacetylated forms **P2**-**P5** of histone H4 peptide $^4$GKGGKGLGKGGAKR$^{17}$, for which both of these methods have so far been unsuccessful [see e.g. Supplementary Table 3 in reference 19].

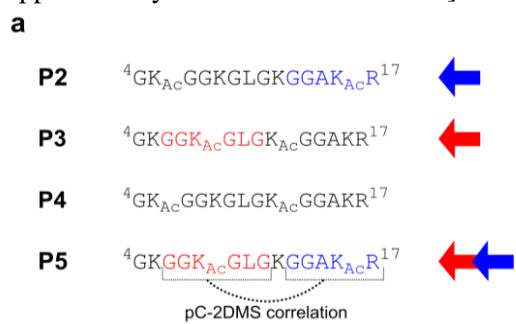
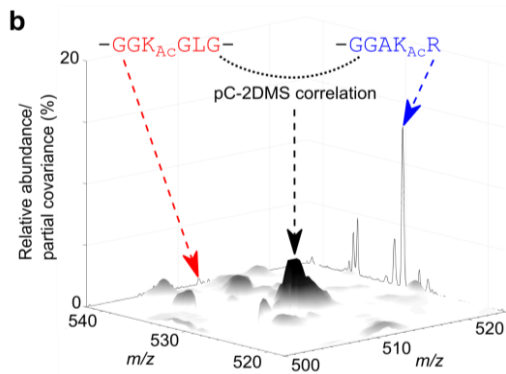
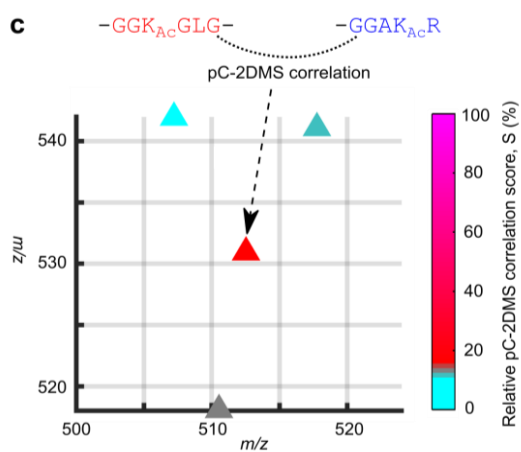

**Figure 3 | The concept of pC-2DMS marker ion correlations. a,** In an arbitrary mixture of fragmented diacetylated isomers **P2-P5** ($K_{Ac}$ = $N^\varepsilon$-acetyllysine), the Lys-8, Lys-16 diacetylated peptide **P5** has no unique 1D marker fragments. For example, the red and blue fragments of **P5** can each be generated by more than one co-fragmented isomer. However, the correlation between these non-unique fragments is unique to **P5** and represents one of the *marker ion correlations* of the peptide. **b,** 3D view of ($m/z$ 500–524) × ($m/z$ 518–542) region of the pC-2DMS map of a mixture of triply protonated **P2–P5,** exhibiting the marker ion correlation of **P5** (all measured marker ion correlations can be found in Supplementary Fig. 5a). The corresponding segments of the averaged 1D CID spectrum, featuring the individual non-unique fragments, are plotted against the back walls of the map. **c,** 2D view of the pC-2DMS map region shown in panel **b** with features extracted using the $S(X,Y)$ correlation score, Eq. 4, (scores are colour-coded).

The correlation-based nature of pC-2DMS enables one to obtain *unique* 2D marker ion correlations between the *non-unique* 1D peptide fragments, solving the formidable challenge of combinatorial modification analysis. This general principle is illustrated in Fig. 3 for Lys-8, Lys-16 diacetylated peptide **P5**[39]. In an arbitrary mixture with three other diacetylated forms of the same amino acid sequence (see **P2**–**P4** in Fig. 3a) **P5** is unable to produce, regardless of fragmentation method or instrumental mass accuracy, any possible unique fragment (either terminal or internal) that would distinguish it from the co-fragmented isomers (see Supplementary Tables 1 and 2). For example, the peptide fragments [GGAK$_{Ac}$R]$^+$ ($y_5^+$, blue) and [GGK$_{Ac}$GLG]$^+$ ($b_{i(3-8)}^+$, red) can each be generated by more than one co-fragmented isomer (marked by arrows of corresponding colour in Fig. 3a). In contrast, the pC-2DMS correlation between the two *non-unique* fragments (Fig. 2b) is *unique* to the modification pattern of **P5** and therefore allows one to unambiguously reveal the occurrence of this structural isomer. We call such pC-2DMS signals isomer-specific *marker ion correlations*. All measured pC-2DMS marker ion correlations of each peptide **P2**–**P5** are given in Supplementary Fig. 5a. Figure 4 demonstrates the use of these *marker ion correlations* to straightforwardly resolve different two- to four-fold mixtures of the peptides **P2**-**P5**, which have proven inaccessible to 1D MS/MS[19,27]. We have also validated the applicability of pC-2DMS to quantitative analysis predicted by Eq. 3 on the four-fold mixture of peptides, see Supplementary Information, Fig. 5b.

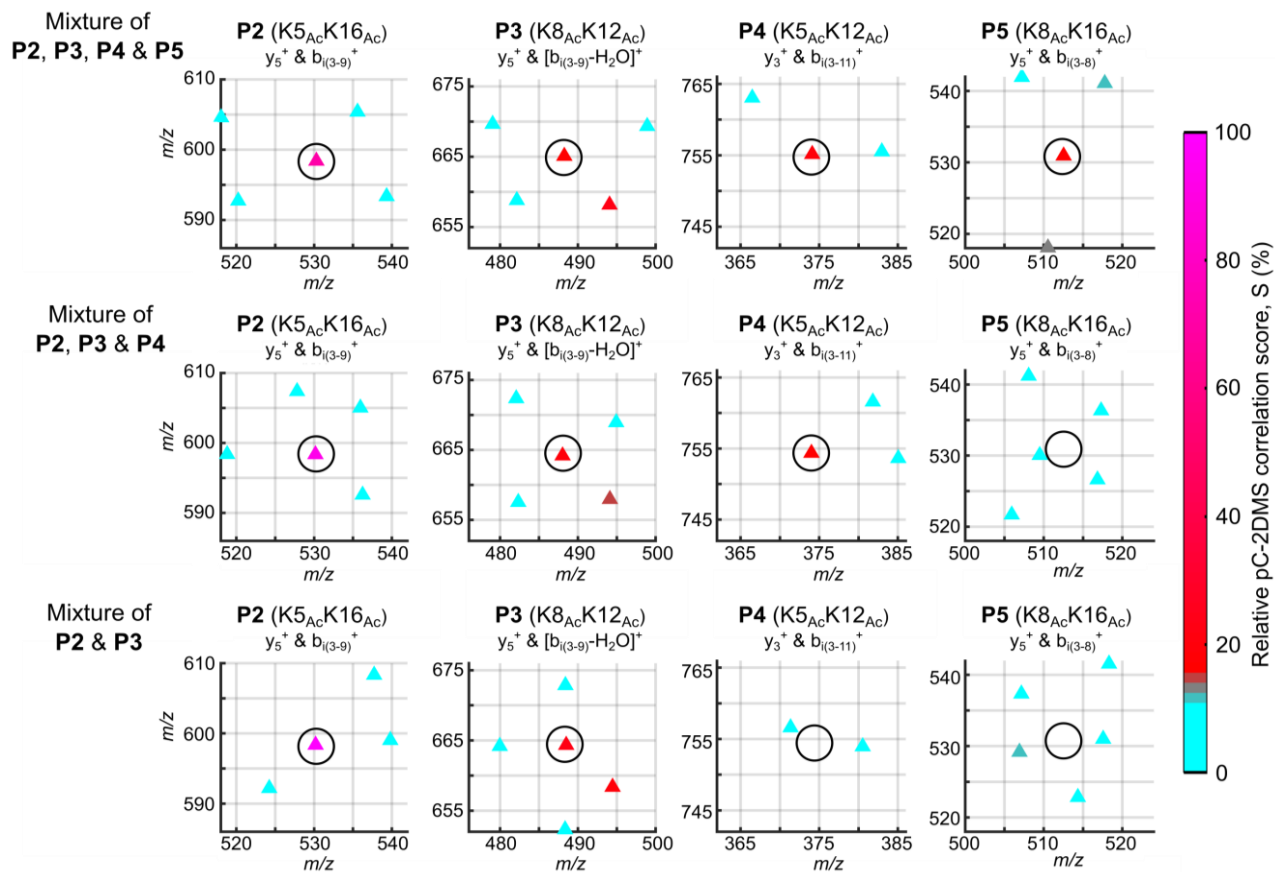

**Figure 4 | Determination of combinatorial modification patterns of unresolved peptide isomers using pC-2DMS.** Three different mixtures of the modified isomers were subjected to pC-2DMS analysis: **P2**, **P3**, **P4** & **P5** (top), **P2**, **P3** & **P4** (middle) and **P2** & **P3** (bottom). pC-2DMS straightforwardly reveals the presence or absence of each of the structural isomers through its isomer-specific marker ion correlations. Only one of the measured marker ion correlations is shown for each isomer (all the measured marker ion correlations for each peptide **P2**-**P5** are shown in Supplementary Fig. 5a).

## Conclusions

In conclusion, we have developed a new two-dimensional mass spectrometry for structural analysis, pC-2DMS, which provides a unique kind of structural information: *fragment-fragment* correlations. These correlations are much more sequence-specific than the individual fragment ion mass-to-charge ratios on which standard 1D MS is based. By matching possible biomolecular sequences to correlated fragment pairs rather than individual fragment ions, we can confidently single out the correct primary structure where the standard approach fails. This new technology has been developed and demonstrated using a commercial benchtop mass spectrometer, with the readily accessible total ion count as the single partial covariance parameter, enabling its immediate wide utilisation as a practical tool. Since the new technique relies on the physical principle of partial covariance mapping, its applicability depends primarily on sufficiently high fragmentation-to-detection efficiency of a mass spectrometer. Once this is ensured, pC-2DMS can be applied to any chemical species under any ionisation and activation methods.

## Methods

Synthetic peptides were prepared by Fmoc strategy for solid-phase synthesis on a multiple automated synthesizer (Syro II, Multisyntech). The synthesis was carried out on preloaded Wang resins. The synthetic peptides were dissolved in a solution of 50% acetonitrile/2% formic acid in water to give concentrations of ~1 μM to ~10 μM for 1D CID acquisitions and 100 fM to ~ 1 μM for pC-2DMS measurements. All mass spectral measurements were performed on a LTQ XL (Thermo Fisher Scientific) linear ion trap mass spectrometer. The samples were infused into the mass spectrometer via a Harvard Apparatus 11 Plus Single Syringe Pump coupled to a Nanospray II Ion Source (Thermo Fisher Scientific) at flow rates of 3-5 µl/min and spray voltages of 1.8 – 2.2 kV, with no flow of auxiliary desolvation gas. The temperature of the ion transfer capillary was held constant at 200°C. The peptide ion of interest was isolated in the linear ion trap and fragmented by collisional-induced dissociation at normalised collision energy of 20% or 35%, activation time of 30 ms and Mathieu q-value of 0.25. The MS/MS scans were performed at scan rates of 16 666 Da/s and 125 000 Da/s. No averaging over microscans was performed in the pC-2DMS experiments.

**Data availability**

The data presented in this study and the codes are available from the corresponding author upon request.

**Acknowledgments**: M. E. A. acknowledges support of the Wellcome Trust through the research fellowship, award WT100093MA. Support was provided by EPSRC programme grant EP/I032517/1, EPSRC/DSTL MURI EP/N018680/1 and Pathways to Impact grant EP/K503733/1, and ERC ASTEX project 290467.

**Author Contributions:** M. E. A. conceived, designed and guided the research. T. D. performed the research. R. A. contributed to the oligonucleotide part of the research. B. C. and S. P. constructed the theoretical model of TIC partial covariance. V. A. guided T. D. in the construction of the automatic pC-2DMS search engine. N. B. assisted in implementing the pC-2DMS search engine. L. J. F. guided T. D. in the development of the peak scoring procedure. R. P. carried out the peptide synthesis. All the authors contributed to the research development, data analysis and manuscript preparation.

**Competing interests:** The authors declare no competing interests.

**Correspondence and requests for materials** should be addressed to M. E. A.

# Supplementary Information

## 1. Formal theory of covariance and partial covariance mapping for noisy Poisson processes

Mikosch and Patchkovskii[34] developed a formal theory of covariance for noisy Poisson processes. For such processes the event rate follows a so-called 'augmented Poisson distribution', which is a Poisson distribution for which the mean event rate $\nu = \gamma \nu_0$ is itself sampled from a normal distribution, with constant $\nu_0$ the central event rate and $\gamma$ centred around $\gamma = 1$. The statistics are analysed in terms of the "compound outcomes", $\mathbb{N}$. For each of the $m$ possible outcomes of the elementary fragmentation process, the $m$-th element of the vector $\mathbb{N}$ gives the number of times this elementary outcome has occurred in the overall, compound outcome. The moments $M(\mathbb{K})$ of the distribution of the compound outcomes are given by a recursive expression[34]:

$$M(\mathbb{K} + \mathbb{I}_j) = P_j \nu_0 M(\mathbb{K}) + P_j \frac{\partial}{\partial P_j} M(\mathbb{K}) + \nu_0^2 \sigma^2 P_j \sum_{l=1}^{m} P_l \sum_{b_l=0}^{k_l-1} \binom{k_l}{b_l} \times M(\mathbb{K} + \mathbb{I}_l(b_l - k_l)) \quad (S1)$$

where $\mathbb{K}$ is the vector $k_1 \ldots k_m$ enumerating the desired moment order, $\mathbb{I}_j$ is a vector of length $m$ with 1 at position j and zeroes elsewhere, $P_j$ is the probability of outcome $j$ for an elementary event, $\sigma$ is the standard deviation of the normal distribution from which $\gamma$ is sampled and $\binom{k_l}{b_l}$ is the multinomial coefficient. The recursion in equation (S1) is initialised by the zeroth moment of a probability distribution (i.e. 1).

This framework can be used to analyse a covariance measurement on a system with known outcome probabilities for the mechanism being probed and known finite detection efficiency. By systematically considering the respective probabilities of each relevant spectral measurement, one can derive an analytical expression for the simple covariance between any number of spectral channels.

Here we have extended used the theoretical approach of Ref. 34 to derive an expression for the partial covariance between two spectral signals where the total number of measured ions (total ion count, TIC) is used as a single partial covariance parameter [Eq. (2)]. We are interested in the partial covariance between two fragment ions X and Y for the case of the induced fragmentation of a biomolecular ion. The total number of parent ions subjected to analysis at each scan is assumed to follow the noise-augmented Poisson distribution, which we have experimentally confirmed in Supplementary Fig. 3. We need to consider the following probabilities for the decomposition of a single parent ion:

1) X and Y are created in the parent ion decomposition, with the probability $P = \alpha_1$
2) X and another molecular fragment Z that is not Y are created in the parent ion decomposition (e.g. Z is the product of a neutral loss or charged loss from Y), $P = \alpha_2$
3) Y and another molecular fragment Z that is not X are created in the parent ion decomposition (e.g. Z is the product of a neutral loss or charged loss from X), $P = \alpha_3$
4) Any two molecular fragments, neither of which are X or Y, are created in the parent ion decomposition (e.g. parent ion dissociates to two other complementary fragment ions A+B), $P = \alpha_4$
5) The parent ion does not fragment, $P = \alpha_5$

We then account for the finite detection efficiency, $\beta$, by considering possible measurement outcomes following an elementary fragmentation event. The following elementary measurement events are relevant for the TIC-based partial covariance:

1) X and Y are both measured at the detector, $P_1 = \alpha_1 \beta^2$
2) Only X is detected (i.e. Y or Z is also produced, but are not observed due to the finite detection efficiency), $P_2 = (\alpha_1 + \alpha_2)\beta(1 - \beta)$
3) Only Y is detected (i.e. X or Z is also produced, but is not observed), $P_3 = (\alpha_1 + \alpha_3)\beta(1 - \beta)$
4) X and another ion Z that is not Y (e.g. Z is the product of a neutral loss from Y) are both measured at the detector, $P_4 = \alpha_2 \beta^2$
5) Y and another ion Z that is not X (e.g. Z is the product of a neutral loss from X) are both measured at the detector, $P_5 = \alpha_3 \beta^2$

6) Only one ion, that is not X or Y, is detected (i.e. parent ion dissociates to two different fragment ions A+B and only A or B is observed, a non-fragmented parent ion is detected, or fragmentation leads to X/Y+Z, but X/Y is not detected), $P_6 = \beta(\alpha_5 + (1-\beta)(\alpha_2 + \alpha_3 + 2\alpha_4))$
7) Any two ions, neither of which are X or Y, are measured at the detector (i.e. parent ion dissociates to two different fragment ions A+B, both of which are measured at the detector), $P_7 = \alpha_4 \beta^2$

Following the procedure outlined in Ref. 34, we obtain the final expression for the TIC-corrected partial covariance between X and Y:

$$pCov(X, Y, TIC) = \beta^2 v_0 \alpha_1 - \frac{\beta(1+\beta)v_0(\alpha_1 + \alpha_2)(\alpha_1 + \alpha_3)}{2(\alpha_s)} + \frac{v_0(\alpha_1 + \alpha_2)(\alpha_1 + \alpha_3)\alpha_5\beta(1 + \beta + v_0\sigma^2\beta(2\alpha_s + \alpha_5(1+\beta)))}{2\alpha_s(2\alpha_s + \alpha_5 + (2\alpha_s + v_0\sigma^2(2\alpha_s + \alpha_5)^2)\beta)} \quad (S2)$$

where $\alpha_s = \alpha_1 + \alpha_2 + \alpha_3 + \alpha_4$. We observe complete parent ion fragmentation in the experiments performed in this work, meaning $\alpha_5 \approx 0$ and the expression simplifies to TIC partial covariance between two fragment ions X and Y as given by:

$$pCov(X, Y, TIC) = \beta^2 v_0 \alpha_1 - \omega v_0,$$
$$\omega = \frac{\beta(1+\beta)(\alpha_1 + \alpha_2)(\alpha_1 + \alpha_3)}{2(\alpha_1 + \alpha_2 + \alpha_3 + \alpha_4)} \quad (S3)$$

where $v_0$ is the average number of trapped ions, $\alpha_1$ is the probability that one of the parent ions under analysis will dissociate to produce X and Y, $\alpha_2$ is the probability that a parent ion dissociates to give X and another fragment ion that is not Y (e.g. there is a neutral loss from Y), $\alpha_3$ is the probability that a parent ion dissociates to give Y and another fragment ion that is not X (e.g. neutral loss from X) and $\alpha_4$ is the probability that a parent ion dissociates to give any two other fragment ions, neither of which are X or Y. $\beta$ is the detection efficiency of the detector, typically required to be ~30% or higher for successful covariance analysis[32].

Equation (S3) shows that the TIC partial covariance between X and Y has a positive contribution that is directly proportional to the probability of a parent ion producing the fragments X and Y, and a negative contribution stemming from competing processes. Given high enough detection efficiency and provided the number of dissociation pathways available to the fragmenting molecule is large enough such that $\alpha_4 \gg \alpha_{1,2,3}$ (practically, this is the case for all fragmenting biological polymers), the first term dominates and the TIC partial covariance is linearly dependent on parent ion abundance and the branching ratio of the process that produces X and Y together. The formal analysis of the TIC-based partial covariance shows that it is false-positive-free: if the TIC partial covariance between a fragment pair is positive, to a desired degree of statistical significance, this pair is known to have been produced in the same fragmentation process, to the same degree of statistical significance.

## 2. Identification of fragment charge state using mass conservation lines

The canonical method for identification of fragment ion charge state in 1D MS is to exploit the small natural abundance of heavier atomic isotopes (e.g. $^{13}C$ at ~1.1% and $^{15}N$ at ~0.4%) to determine the charge of a fragment ion, by measuring the *m/z* difference between its isotopic envelope signals. Consecutive isotopic peaks are separated in *m/z* by $\frac{1}{z}$, where z is the charge of the fragment ion. To correctly infer the charge state of the fragment ion from the isotopic envelope, an accurate and well-resolved measurement of the *m/z* difference between the peaks in the isotopic envelope is required.

The pC-2DMS presents a novel way to extract the charge states of a pair of fragment ions – through the gradient of the mass conservation line upon which their correlation falls. If the two correlated ions measured at $m_1/z_1$ & $m_2/z_2$ are the intact primary products of the dissociation of a parent ion with mass Mr$_{Par}$, then the sum of their masses must be equal to the mass of the parent ion that dissociated to produce them:

$$(m_1/z_1) \times z_1 + (m_2/z_2) \times z_2 = Mr_{Par}. \tag{S4}$$

This means that any such correlation, located on a pC-2DMS map at x=$m_1/z_1$, y=$m_2/z_2$, will fall along a straight line of the form:

$$x \times z_1 + y \times z_2 = Mr_{Par}. \tag{S5}$$

This can be rearranged to:

$$y = -\frac{z_1}{z_2} \times x + \frac{Mr_{Par}}{z_2}, \tag{S6}$$

which defines a straight line with gradient $-\frac{z_1}{z_2}$ and y-intercept $\frac{Mr_{Par}}{z_2}$. Therefore, if a set of correlation islands lying on a particular mass conservation line is identified, both the ratio between the charge states $\frac{z_1}{z_2}$ and the value of $Mr_{Par}$ can be identified as well. Provided even a limited knowledge on the charge state of the parent ion (e.g. a broad range of potential charge states), this enables the charge of each correlated fragment ion, as well as the exact charge and mass of the parent ion responsible for the mass conservation line, to be calculated.

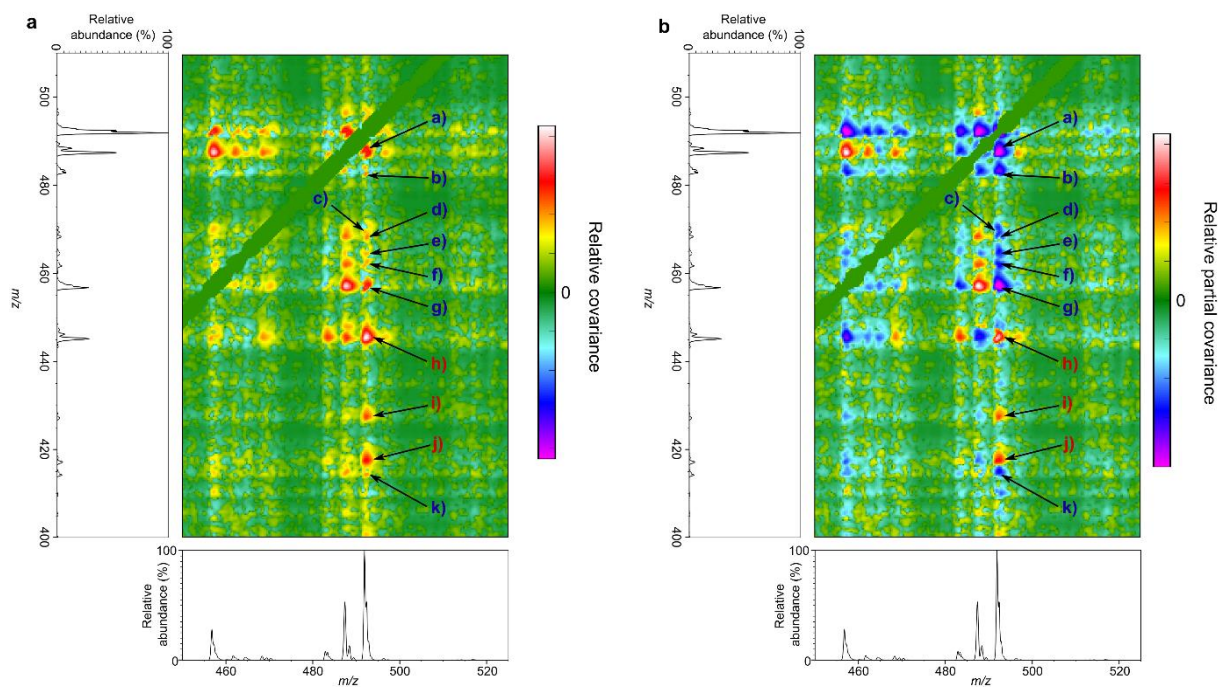

**Supplementary Figure 1 | Simple covariance mapping *vs.* TIC partial covariance mapping. a,** 2D view of a ($m/z$ 450–525) × ($m/z$ 400–510) region of the 2D *simple* covariance map of CID on triply protonated peptide **P1** (VTIMPK$_{Ac}$DIQLAR), constructed using equation (1). The simple covariance map exhibits positive covariance between all spectral signals as a result of fluctuating experimental parameters causing all signal intensities to rise and fall synchronously. The corresponding region of the 1D CID spectrum is shown along the *x*- and *y*-axes. **b,** 2D view of the same region of the TIC *partial* covariance pC-2DMS map of peptide **P1**, constructed using equation (2). The pC-2DMS map reveals the peptide fragments of common origin (true correlations appearing as positive red/yellow islands on the pC-2DMS map) whose signal intensities fluctuate together as a result of them being created along the same fragmentation pathway, whilst the extrinsic correlations are suppressed (negative blue/magenta islands). The full (non-truncated) pC-2DMS maps are symmetric with respect to the diagonal ($x = y$), removed for clarity, along which each spectral signal is trivially correlated with itself. Both true (red letters) and extrinsic (blue letters) correlation islands are annotated for the major signals corresponding to correlations of the peptide fragment $y_8^{2+}$: a) $y_8^{2+}$ & $y_4^+$, b) $y_8^{2+}$ & $[y_8-H_2O]^{2+}$, c) $y_8^{2+}$ & $b_8^{2+}$, d) $y_8^{2+}$ & $a_{i(5-8)}^+$, e) $y_8^{2+}$ & $[M+3H-2H_2O]^{3+}$, f) $y_8^{2+}$ & $[b_8-H_2O]^{2+}$, g) $y_8^{2+}$ & $a_8^{2+}$, h) $y_8^{2+}$ & $b_4^+$, i) $y_8^{2+}$ & $[b_4-H_2O]^+$, j) $y_8^{2+}$ & $a_4^+$, k) $y_8^{2+}$ & $[y_1-2NH_3]^+$, where M is the neutral peptide.

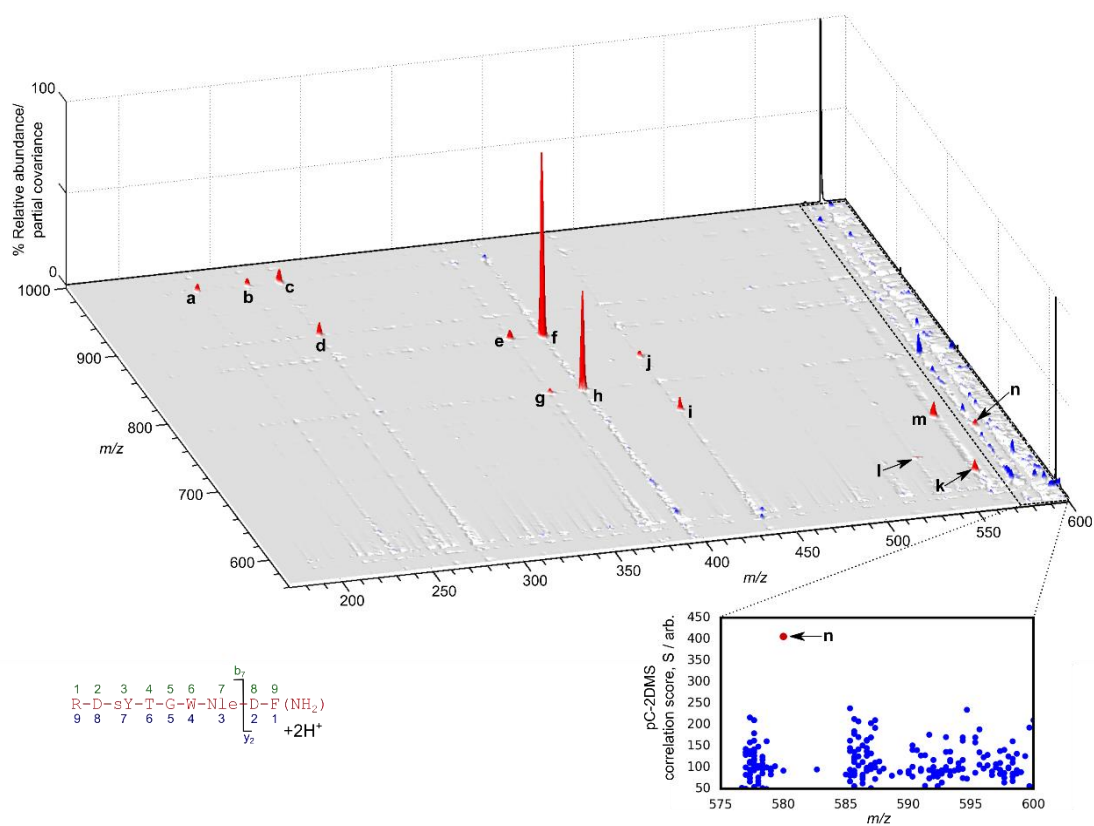

**Supplementary Figure 2 | Identification of low-intensity true pC-2DMS signals by jackknife resampling.** Off-diagonal ($m/z$ 170–600) × ($m/z$ 560–1000) region of the pC-2DMS map of doubly protonated sulphopeptide RDY(SO$_3$H)TGW-Nle-DF-NH$_2$. The vertical axis of the map represents the value of the TIC partial covariance [Eq. (2)] and the horizontal axes display the mass-to-charge ratio ($m/z$). The corresponding parts of the averaged 1D CID spectrum, dominated by the [M+2H-SO$_3$]$^{2+}$ signal and as a result revealing the strong suppression of peptide sequence-specific fragments, are plotted against the back walls of the map. The main annotated pC-2DMS signals (red) are: **a)** [y$_2$-HCONH$_2$]$^+$ & b$_7$$^+$, **b)** [y$_2$-NH$_3$]$^+$ & b$_7$$^+$, **c)** y$_2$$^+$ & b$_7$$^+$, **d)** y$_2$$^+$ & [b$_7$-SO$_3$]$^+$, **e)** [y$_3$-H$_2$O]$^+$ & b$_6$$^+$, **f)** y$_3$$^+$ & b$_6$$^+$, **g)** [b$_6$-SO$_3$]$^+$ & [y$_3$-H$_2$O]$^+$, **h)** [b$_6$-SO$_3$]$^+$ & y$_3$$^+$, **i)** y$_6$$^+$ & [b$_3$-SO$_3$]$^+$, **j)** [y$_3$+HNCNH]$^+$ & [b$_6$-HNCNH]$^+$, **k)** z$_4$$^+$ & [c$_5$-SO$_3$]$^+$, **l)** y$_5$$^+$ & [b$_4$-SO$_3$]$^+$, **m)** z$_4$$^+$ & c$_5$$^+$, **n)** y$_4$$^+$ & b$_5$$^+$. Whilst the majority of the map section presents little statistical noise and the true correlation islands are readily identifiable, the area marked by the dotted line ($m/z$ 575–600) × ($m/z$ 560–1000) is populated by a significant number of scattered strong features (blue) arising due to statistical noise associated with the dominant 1D signals. The inset shows that these blue 2D signals are confidently identified as statistical noise by scoring them using jackknife resampling of the original dataset [equation (4)], while the true correlation peak **n** is readily singled out on the basis of its outstanding pC-2DMS correlation score (red spot). Note, the high sensitivity of pC-2DMS enables to identify intact sulfated fragment ions (**a, b, c, e, f, j, m, n**), greatly assisting in the localisation of the post-translational modification under CID.

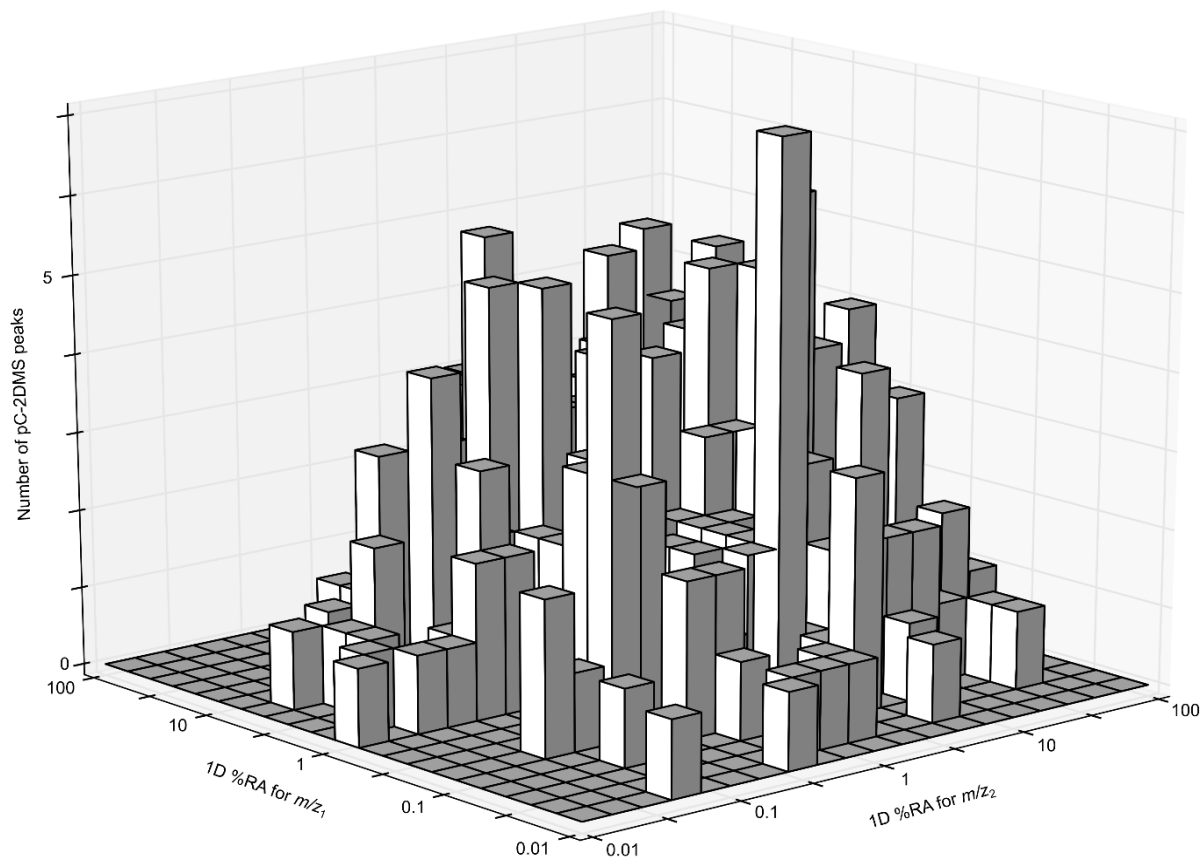

**Supplementary Figure 3 | Distribution of 1D MS/MS relative abundances (RA, %) of fragment ions correlated by pC-2DMS.** The scored lists of pC-2DMS correlations of a representative sample of nine peptide ions** was used to construct the two-dimensional histogram. The number of identified fragment correlation signals ($z$ axis) is plotted as a function of the 1D MS/MS relative abundances of the two correlating fragments, denoted as $(m/z)_1$ and $(m/z)_2$. $(m/z)_1$ is chosen to be the smaller of the two correlated $m/z$ values to avoid double counting. Note the logarithmic bin size. The histogram shows that fragments at very low % RA do not necessarily require a high % RA correlating fragment to produce top-scoring pC-2DMS features. Thus, for example, correlations between fragments of 0.1% RA and 1% RA are very common. The distribution also demonstrates that pC-2DMS is able to simultaneously reveal the relationships of molecular fragments spanning more than three orders of magnitude of % RA in 1D MS/MS.

**[GGNFSGR(Me)GGFGGSR+3H]$^{3+}$ (R(Me) = N$^G$-monomethylarginine), [EQFDDpYGHMRF-NH$_2$+3H]$^{3+}$ (pY = phosphotyrosine), [LGEY(nitro)GFQNAILVR+2H]$^{2+}$ (Y(nitro) = 3-nitrotyrosine), [GWGR(Me$_2$)EENLFSWK+3H]$^{3+}$ (R(Me$_2$) = N$^G$,N$^{G'}$-dimethylarginine), [TWpTLCGTVEY+2H]$^{2+}$ (pT = phosphothreonine), [LGEY(nitro)GFQNAILVR+3H]$^{3+}$ (Y(nitro) = 3-nitrotyrosine), [VTIMPKDIQLAR+3H]$^{3+}$, [SApTPEALAFVR+2H]$^{2+}$ (pT = phosphothreonine), [EQFDDpYGHMRF-NH2+2H]$^{2+}$ (pY = phosphotyrosine).

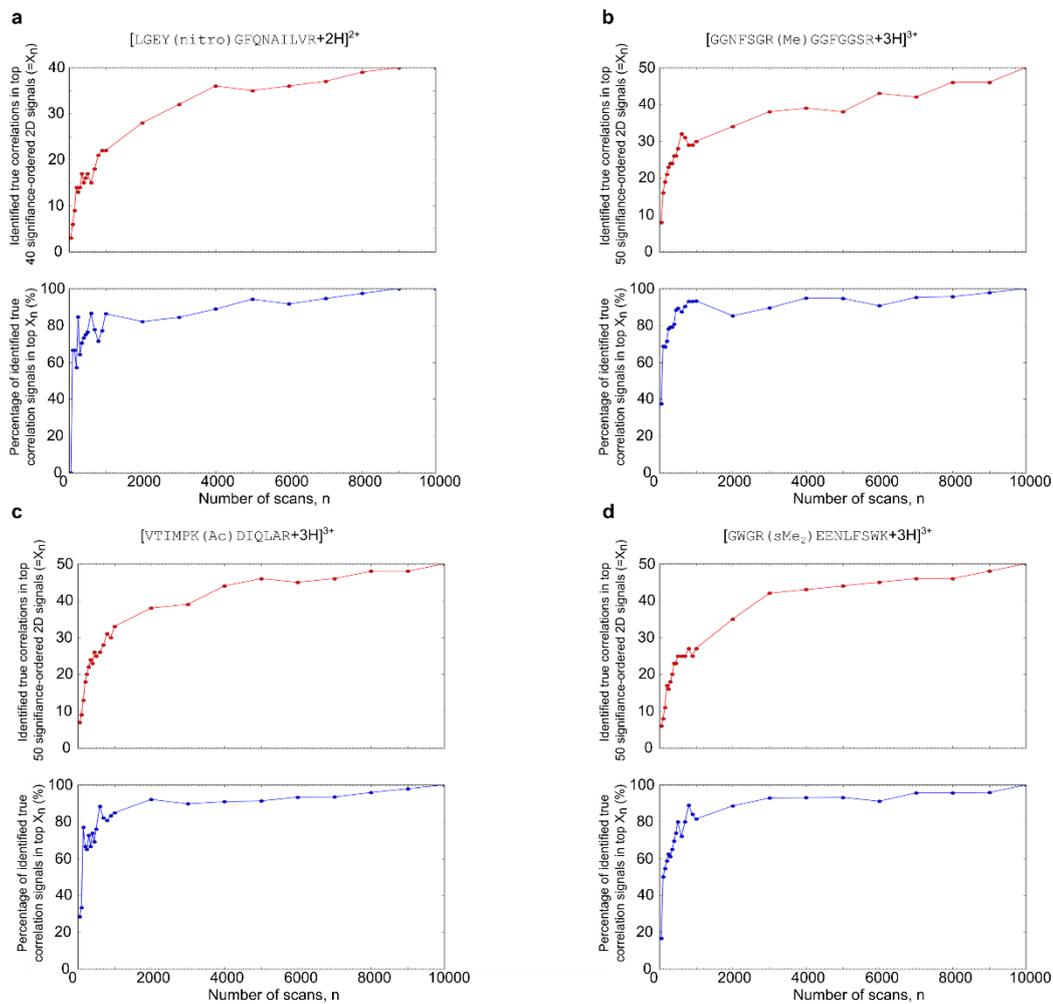

**Supplementary Figure 4 | Dependence of pC-2DMS signals on number of scans acquired.** Panels **a-d** show the convergence of the pC-2DMS correlation score-ranked signal lists with number of scans for a representative sample of doubly and triply charged peptide ions. We choose the manually verified top N (40 for a doubly charged parent ion and 50 for a triply charged parent ion) correlations obtained at 10,000 scans as a reference. The red curves (upper panels) give the number of the top N correlations calculated at each number of scans which are also found in the verified reference correlations (we call this value X), while the blue curve (lower panels) gives the percentage of these manually verified correlations that appear in the X top-scored signals (this gives a measure of the fidelity of the top X signals). The data show that it is possible to extract a very high quality signal list allowing confident sequence assignment after only ~1000 scans.

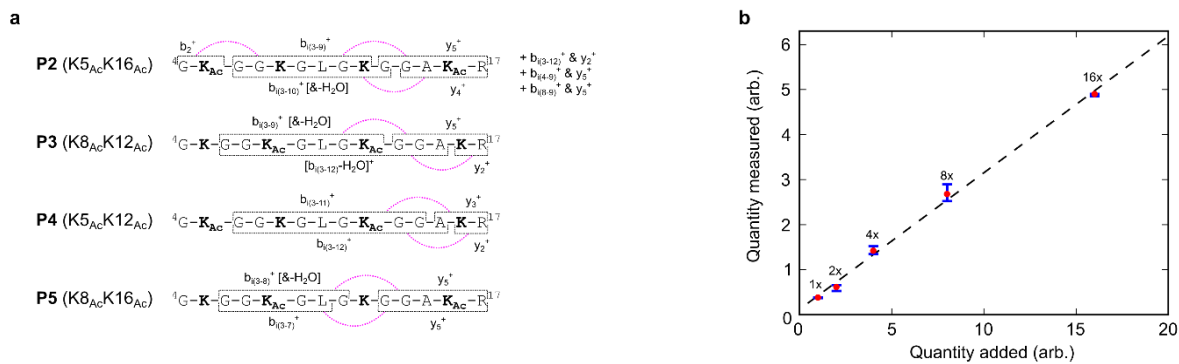

**Supplementary Figure 5 | Resolution of combinatorially modified isomeric peptides using pC-2DMS. a,** the measured unique pC-2DMS signal channels, which enable identification of each of the four positional isomers **P2**–**P5** in an arbitrary mixture of all four, are shown by magenta arcs. **b,** pC-2DMS analysis of five different mixtures of all four isomers **P2**–**P5** with progressively increased relative molar concentration of the isomer **P2**. 'Quantity measured' is the ratio between the volume of the highest scoring partial covariance feature unique to **P2** ($y_{5,Ac}^+$ & $b_{i(3-9)}^+$, found among the top scoring 25 pC-2DMS correlations in each of the five mixtures) and the volume of one of the reference features which cannot be produced by **P2** (found in the top scoring 25 pC-2DMS correlations for all five mixtures): $y_5^{2+}$ & $b_{9,2Ac}^+$, $y_5^+$ & $[b_{9,2Ac}-NH_3]^{2+}$, $y_5^+$ & $[b_{9,2Ac}-C_7H_{11}NO-CO]^+$. The error bars show the dependence of the quantification on the particular choice of the reference correlation, demonstrating consistently accurate results for each of the three chosen references. 'Quantity added' represents the actual relative increase in the **P2** concentration. The dashed line represents the best linear fit. These results demonstrate that the volume of a unique pC-2DMS feature provides a highly accurate measure of the quantity of the measured isomer.

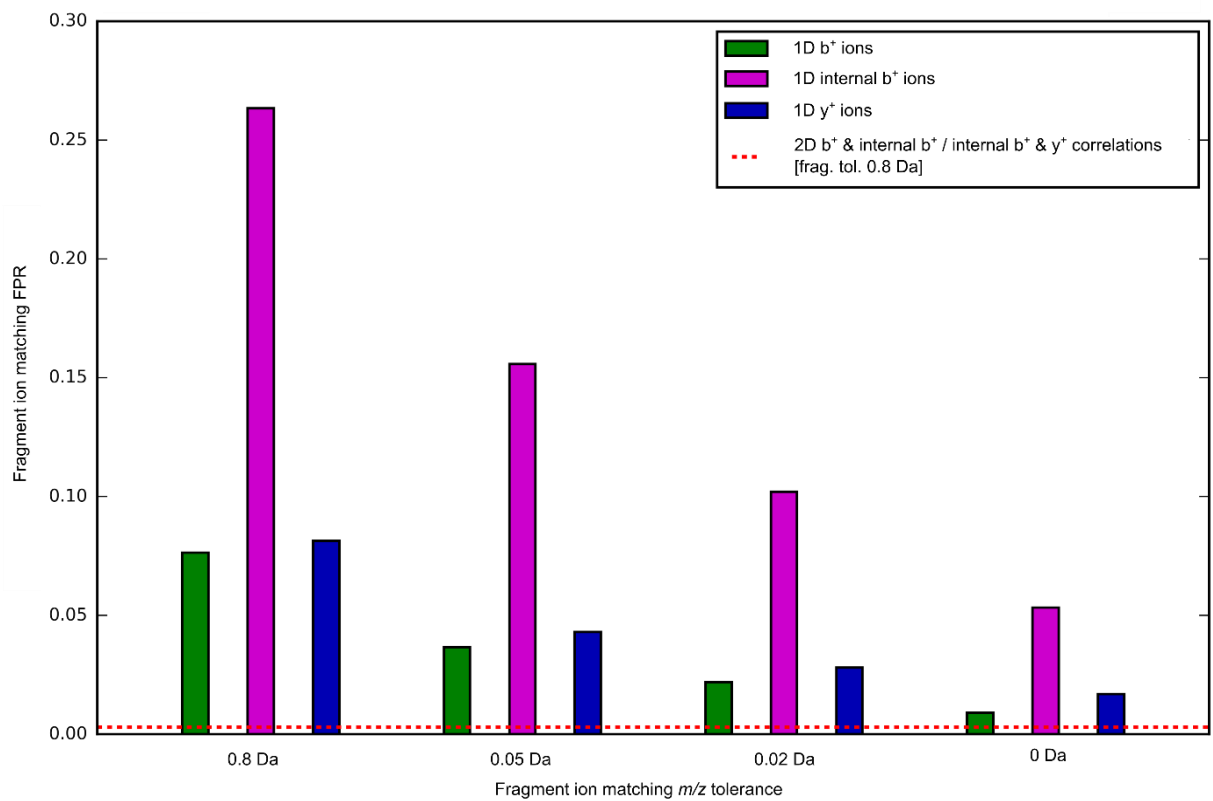

**Supplementary Figure 6 | Reduction in false positive rate for matching pC-2DMS signals *vs.* matching 1D MS/MS signals.** Bars show the estimated false positive rates (FPRs), per candidate database peptide sequence, for matching *m/z* values of individual internal $b^+$- ions (magenta bars) and terminal $b^+$ and $y^+$ ions (green and blue bars, respectively) to *m/z* values of individual fragment ions derived from *in silico* fragmentation of the isobaric (within 7 ppm) peptide sequences of a tryptic digest of the UniProt/Swiss-Prot database. Matching of the individual ions is analysed at fragment ion *m/z* tolerances of 0.8, 0.05, 0.02 and 0 Da (matching isomeric peptide fragments only) and calculated according to equation (4), with all fragment ion lengths between 2 and 15 residues considered. The dashed line shows the estimated FPR for matching $b^+$ & internal $b^+$ and internal $b^+$ & $y^+$ fragment ion pair correlations to all possible theoretical fragment pair correlations of the isobaric (within 7 ppm) database peptide sequences at fragment ion *m/z* tolerance of 0.8 Da only. The data show that by correlating an internal fragment with a corresponding terminal ion, the FPR is dramatically reduced. Even compared to the aggregate FPR for the idealistic limit of matching *individual* $b^+$-ions and $y^+$-ions at infinite *m/z* resolution (0 Da fragment ion tolerance), the aggregate FPR for matching the corresponding internal ion & terminal ion *correlations* at fragment ion tolerance of 0.8 Da (dashed line) is found to be lower by a factor of 4 (0.0124 for 1D $b^+$-ions and $y^+$-ions at 0 Da *vs.* 0.00306 for 2D $b^+$ & internal $b^+$ and internal $b^+$ & $y^+$ correlations at 0.8 Da). Each presented FPR was obtained as an average over all fragment ions/fragment correlations generated from five different sets of 10,000 randomly chosen peptides from an *in silico* tryptic digest of the database. The FPR for matching correlations between complementary terminal $b^+$ & $y^+$-ions was also investigated. As expected on the basis of simple mass conservation arguments and the tight parent ion *m/z* tolerance used for the parent ion pre-selection, the 2D complementary terminal $b^+$ & $y^+$-ion FPR was found, at 0.0388, to be of the same order of magnitude but still substantially lower than the FPR for matching individual $b^+$-ions (0.0764) at *m/z* tolerance of 0.8 Da.

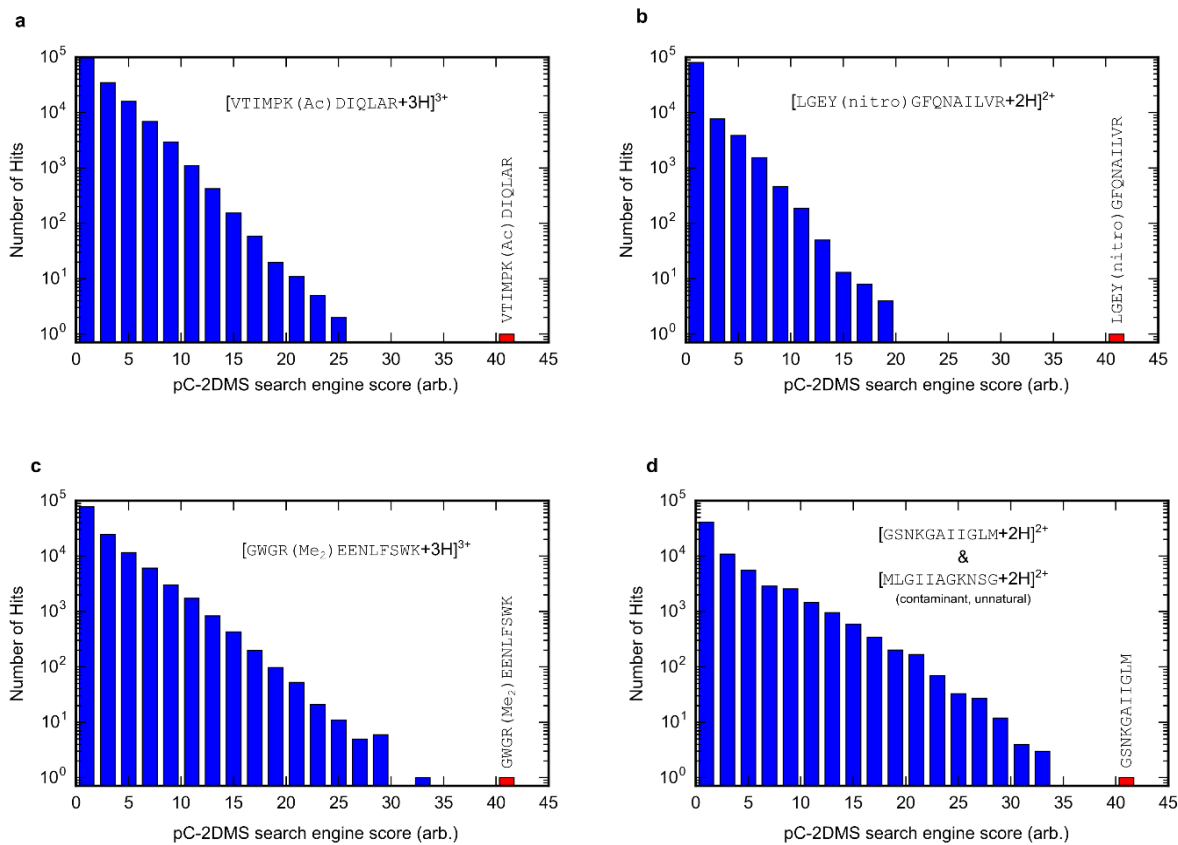

**Supplementary Figure 7 | Spectrum-to-structure matching by prototypical pC-2DMS database search engine.** The histograms **a-d** demonstrate identification of peptide sequences based on their pC-2DMS fragment ion pair correlations only, using our prototypical database search engine. The correct peptides (red bars) are distinguished from a pool of candidate peptides (blue bars) derived from the digest of the UniProt/SwissProt protein database (no enzymatic specificity, 5 ppm peptide $m/z$ tolerance) on the basis of their outstanding pC-2DMS search engine scores. Fragment correlation $m/z$ tolerance was 0.8 Da. Note the logarithmic scale on the $y$-axes of each plot, demonstrating the large number of peptide sequences subjected to correlation matching. **a,** pC-2DMS search engine results for triply protonated peptide VTIMPK(Ac)DIQLAR (Histone H3). Lysine acetylation was selected as a variable modification for the search. **b,** pC-2DMS search engine results for doubly protonated peptide LGEY(nitro)GFQNAILVR (Serum Albumin [Mus Pahari]). Tyrosine nitration was selected as a variable modification for the search. The isomeric database sequences LGEY(nitro)GFQNALLVR and LGEY(nitro)GFQNALIVR, which were trivially awarded the same pC-2DMS search engine score, are not plotted on the histogram. **c,** pC-2DMS search engine results for triply protonated peptide GWGR(Me$_2$)EENLFSWK, R(Me$_2$)=N$^G$,N$^{G'}$-dimethylarginine (Coilin). Arginine dimethylation was specified as variable modification for the search. **d,** pC-2DMS search engine results for identification of the sequence GSNKGAIIGLM (Amyloid Beta peptide) from the pC-2DMS map of an equimolar mixture with its unnatural palindromic reverse isomer MLGIIAGKNSG, introduced to simulate the co-isolation of an abundant contaminant isobaric ion. The high specificity of the pC-2DMS correlation matching (see Supplementary Fig. 6) enables the search engine to uniquely identify the naturally occurring sequence despite a multitude of abundant contaminant correlations from the unnatural isomer, giving it a stand-out pC-2DMS correlation score.

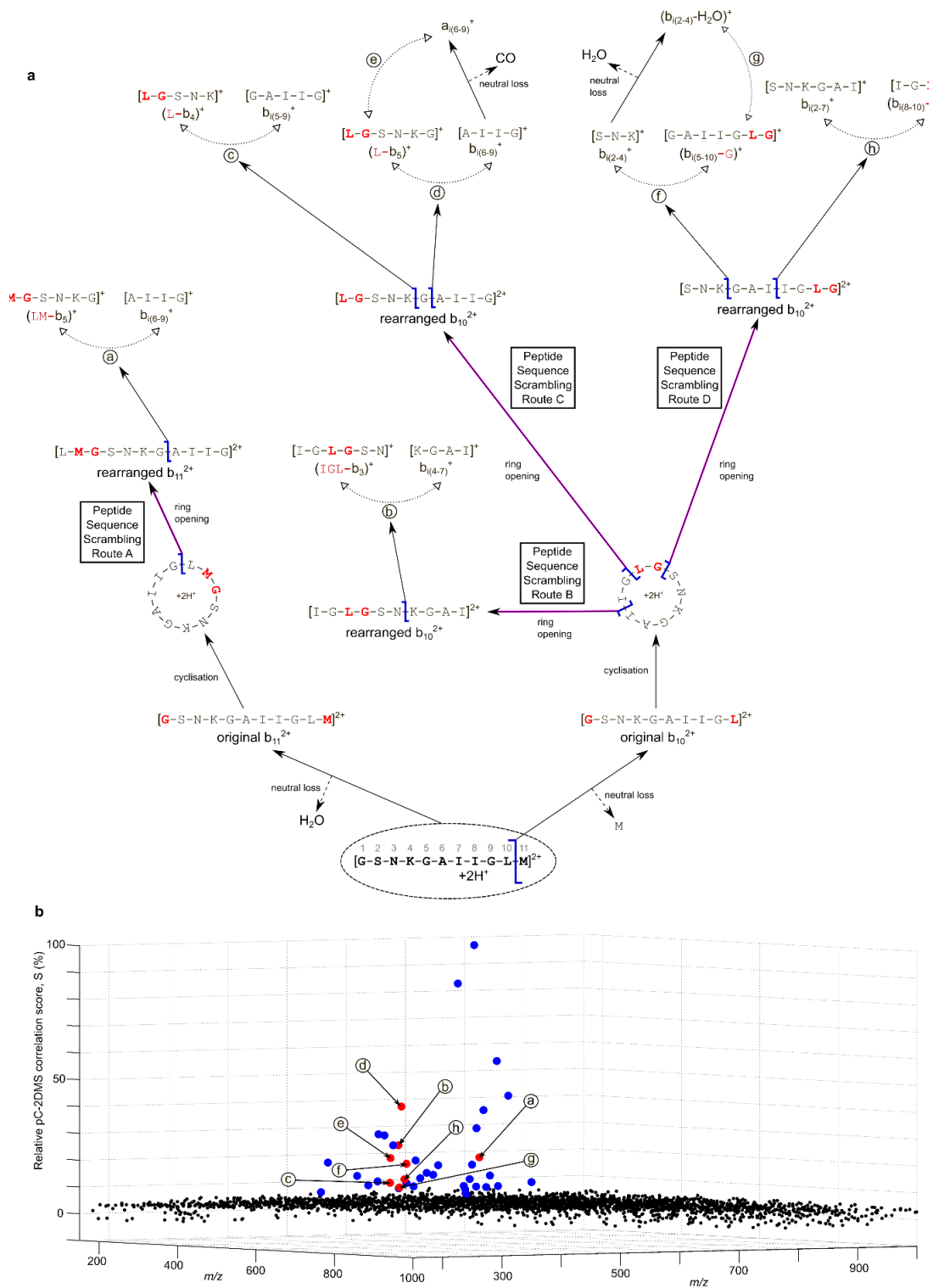

**Supplementary Figure 8 | Formation tree of anomalous fragments of doubly protonated peptide GSNKGAIIGLM (A4 Bovin) reconstructed through pC-2DMS fragment ion correlations. a,** pC-2DMS data reveal fragment ion correlations ⓐ-ⓗ which unambiguously indicate the scrambling of the original peptide sequence within the $b_{10}^{2+}$ and $b_{11}^{2+}$ fragments. The pC-2DMS fragment ion connectivity enables one to establish the cyclisation-based route of the anomalous fragment formation without the need for isomeric model peptide studies[5*]. Terminal residues of the unscrambled peptide fragments are marked in red to illustrate their relocation following the scrambling rearrangement. **b,** three-dimensional scatter plot showing the relative pC-2DMS correlation score $S(X,Y)$ [equation 4], of all resampled pC-2DMS features (*z*-axis), with the two *m/z* values of the correlated fragment ions plotted on the *x*- and *y*-axes. The fragment pair correlations ⓐ-ⓗ shown in panel **a** (red spots) are among the forty highest pC-2DMS correlation-score signals, the rest of which are plotted in blue.

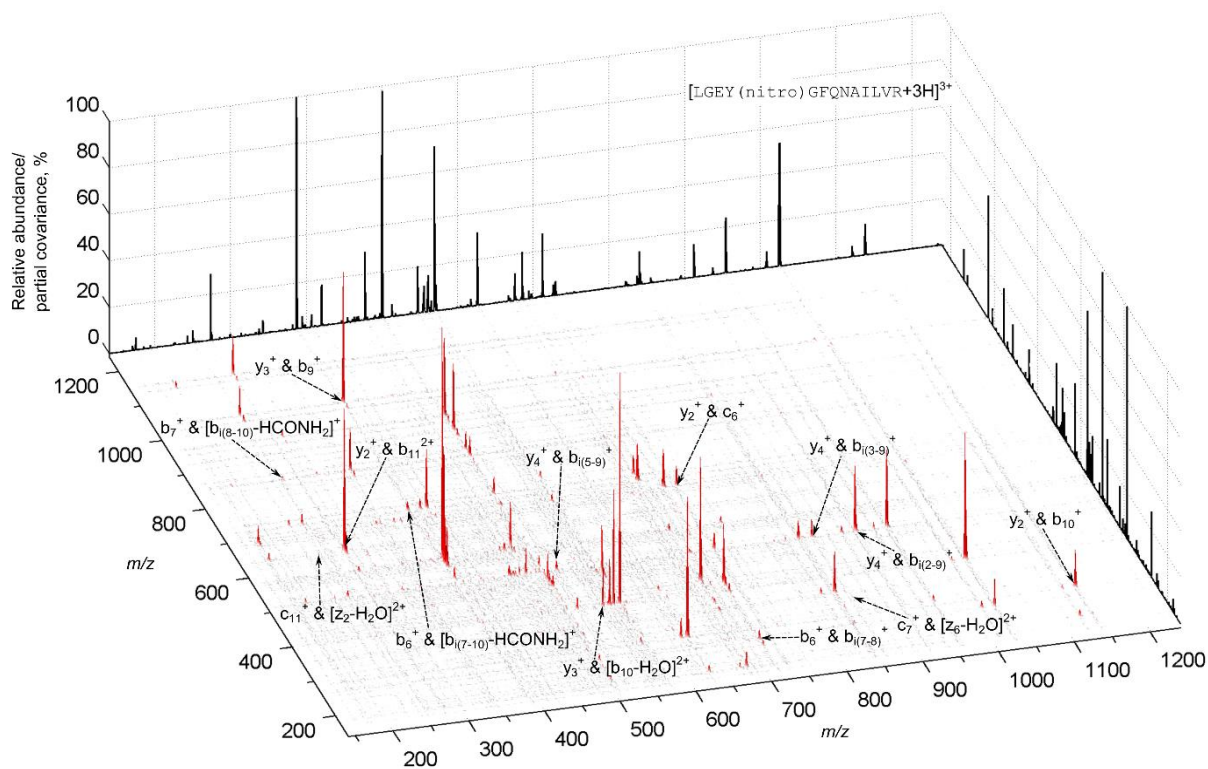

**Supplementary Figure 9 | pC-2DMS map of the triply protonated nitrated peptide LGEY(nitro)GFQNAILVR, Y(nitro)=3-nitrotyrosine**. The $m/z$ values of the correlating fragments are plotted along the $x$- and $y$-axes whilst the $z$-axis represents the TIC partial covariance [Eq. (2)], normalised to the highest off-diagonal pC-2DMS peak. The autocorrelation line, which trivially correlates each spectral signal to itself along the diagonal ($x = y$), has been removed and a standard 8x8 median filtering procedure has been applied to the map for visual clarity. The black line graph plotted against the back walls of the pC-2DMS map is the averaged standard 1D MS/MS mass spectrum. A non-exhaustive selection of top scoring pC-2DMS peaks is labelled, including very weak signals corresponding to the true (intrinsic) fragment ion correlations revealed using pC-2DMS scoring.

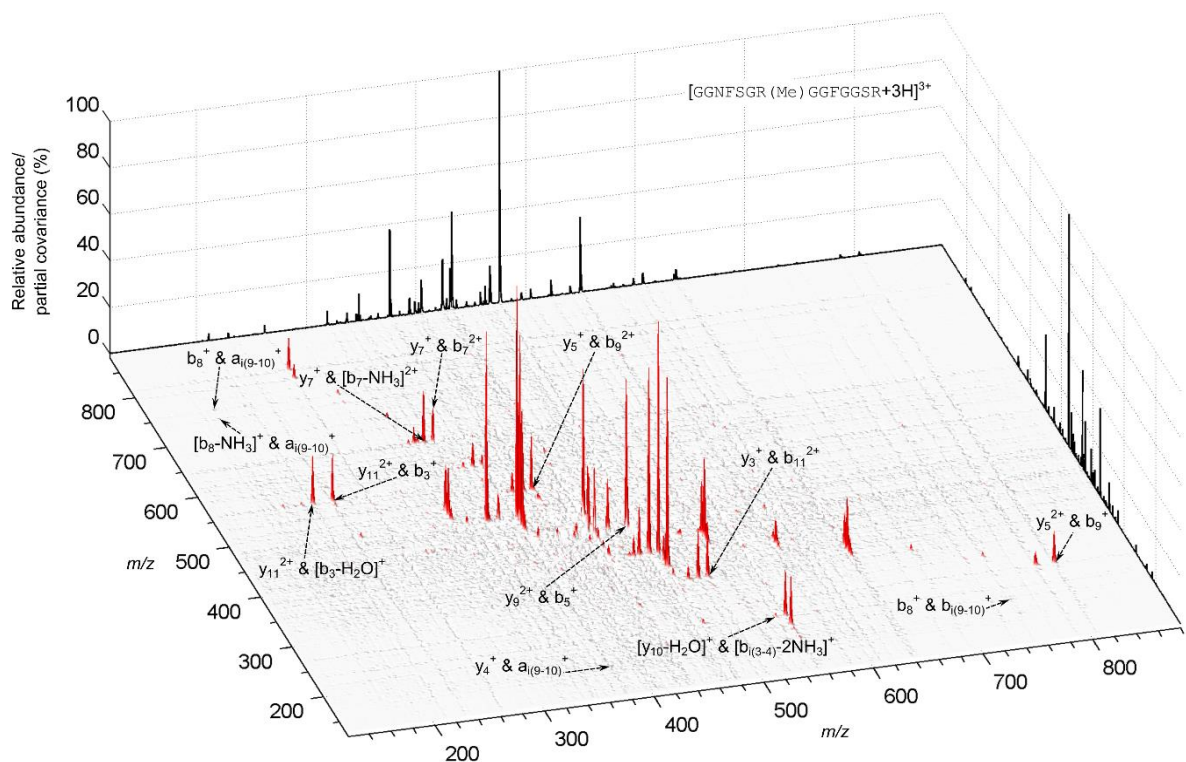

**Supplementary Figure 10 | pC-2DMS map of the triply protonated methylated peptide GGNFSGR(Me)GGFGGSR, R(Me)=N$^G$-monomethylarginine**. Details are the same as for Supplementary Fig. 9.

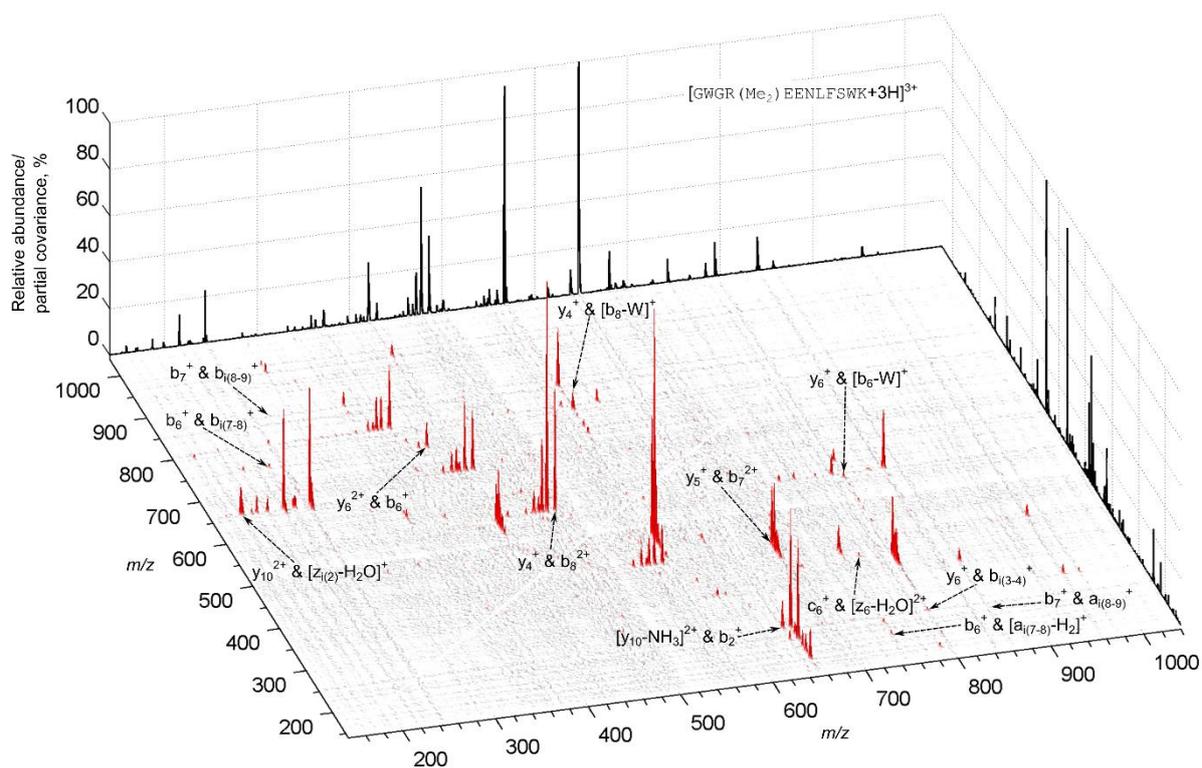

**Supplementary Figure 11 | pC-2DMS map of the triply protonated methylated peptide GWGR(Me$_2$)EENLFSWK, R(Me$_2$)=N$^G$,N$^{G'}$-dimethylarginine**. Details are the same as for Supplementary Fig. 9.

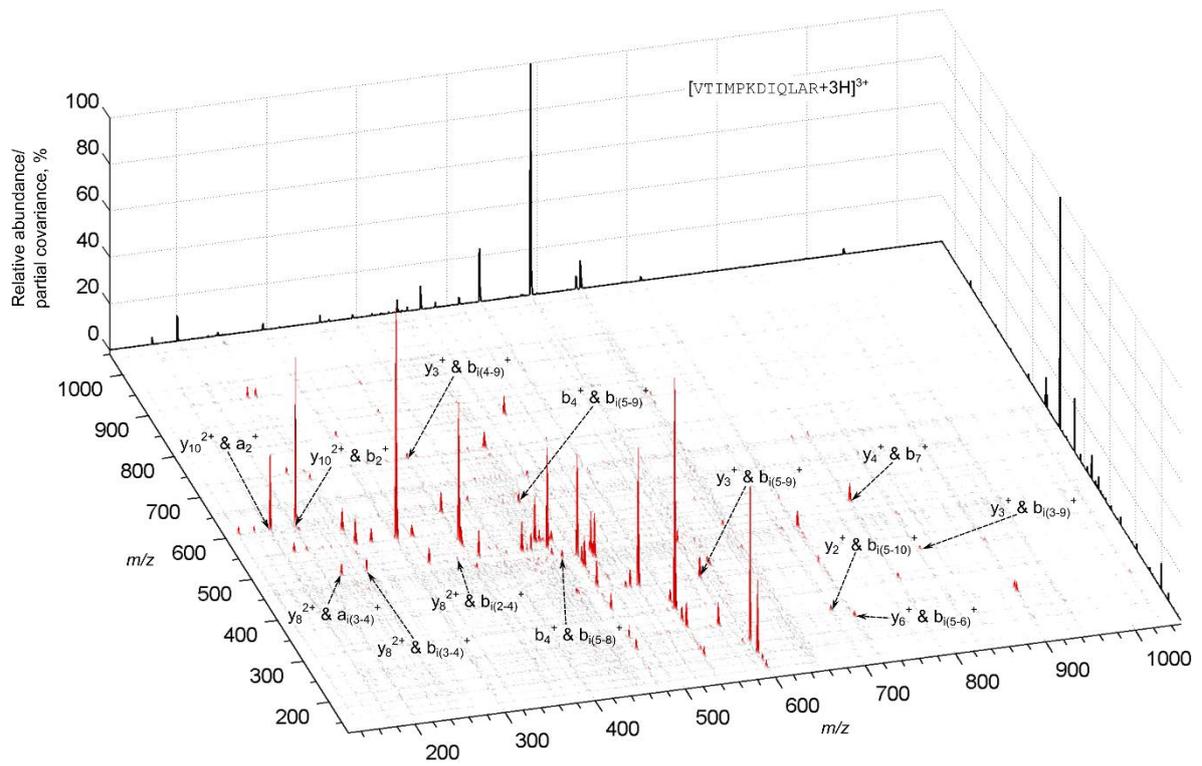

**Supplementary Figure 12 | pC-2DMS map of the triply protonated unmodified peptide VTIMPKDIQLAR.** Details are the same as for Supplementary Fig. 9.

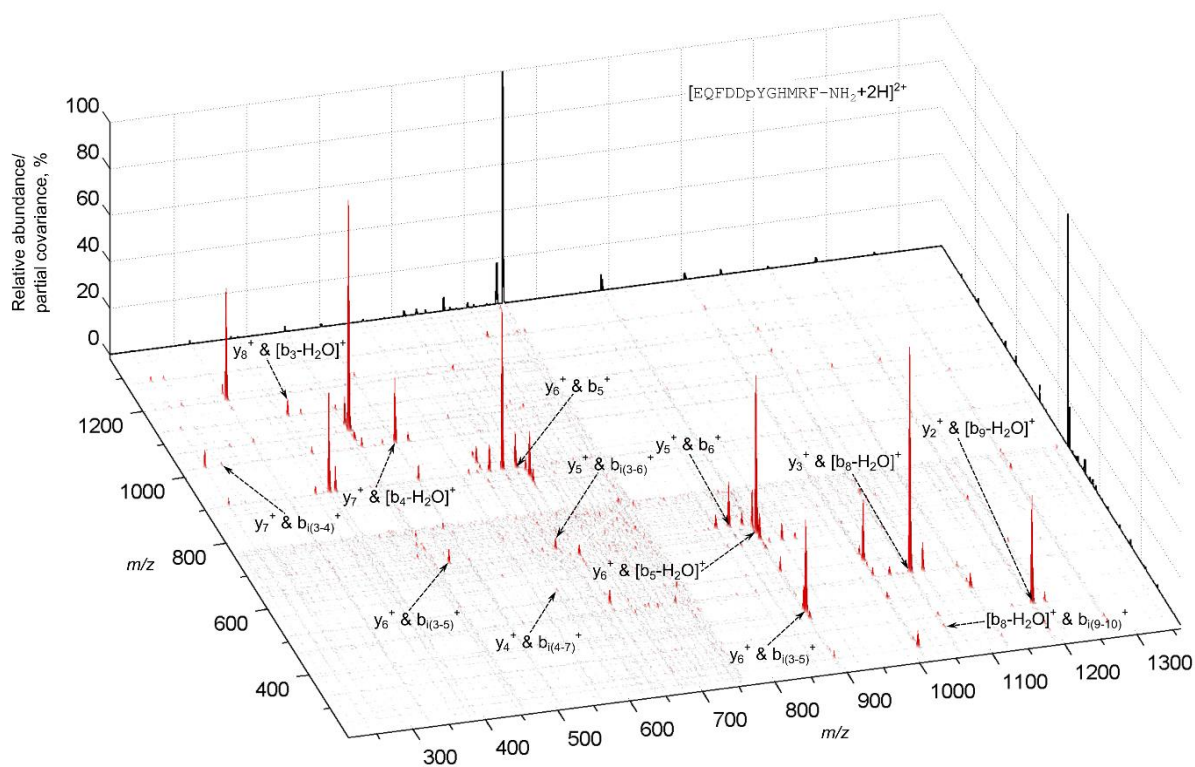

**Supplementary Figure 13 | pC-2DMS map of the doubly protonated phosphorylated peptide EQFDDpYGHMRF-NH$_2$, pY=phosphotyrosine**. Details are the same as for Supplementary Fig. 10. Note that the 1D averaged MS/MS spectrum is dominated by the structurally uninformative signal of the neutral loss of water from the parent ion, which is facilitated by the N-terminal glutamic acid residue of the peptide. Such structurally uninformative neutral loss fragmentations of the parent ions result in only one measurable fragment and thus do not produce any pC-2DMS correlation. It is a general property of pC-2DMS that it can reveal structurally informative signals even when the corresponding 1D spectrum is dominated by parent ion neutral loss (see also e.g. Supplementary Fig. 3).

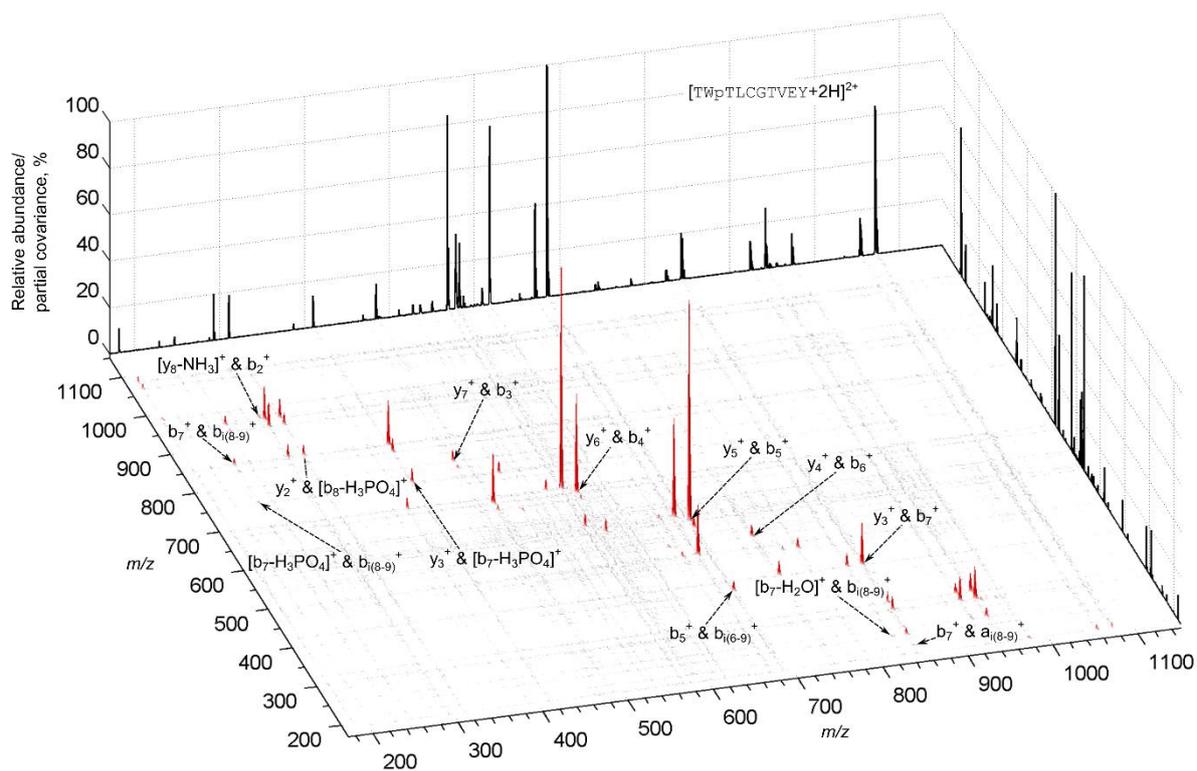

**Supplementary Figure 14 | pC-2DMS map of the doubly protonated phosphorylated peptide TWpTLCGTVEY, pT=phosphothreonine**. Details are the same as for Supplementary Fig. 9.

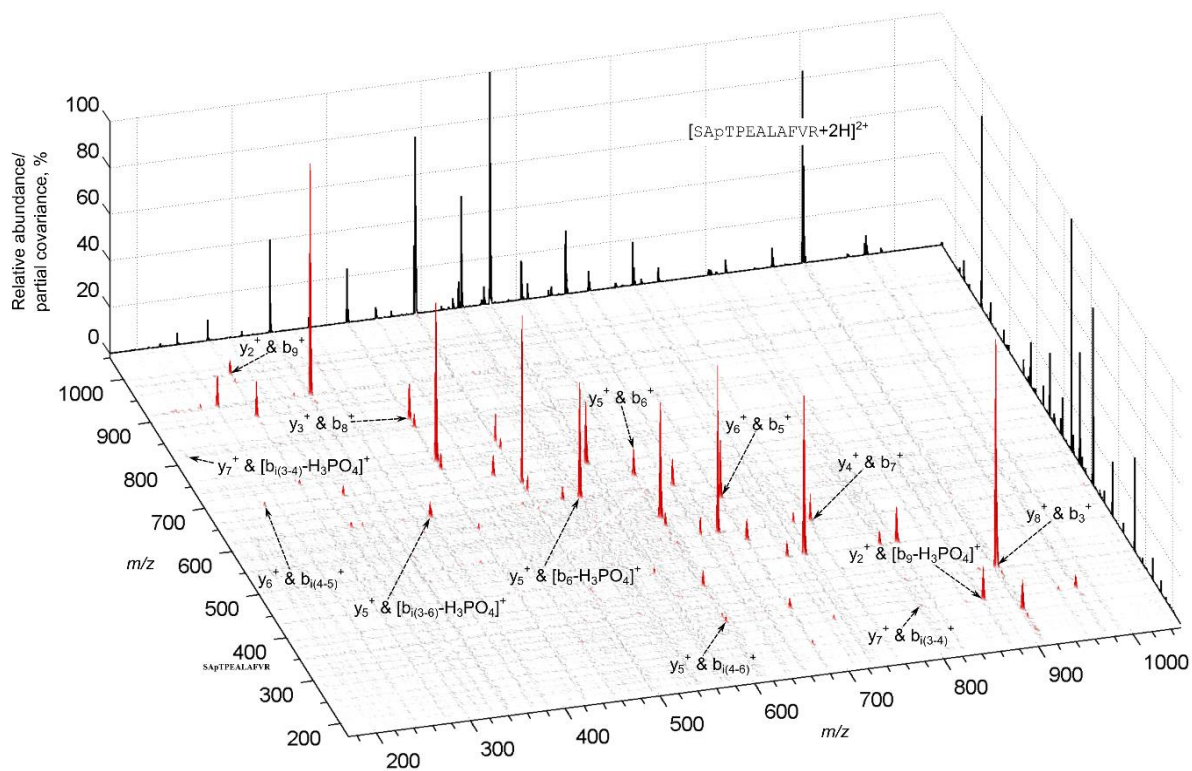

**Supplementary Figure 17 | pC-2DMS map of the doubly protonated phosphorylated peptide SApTPEALAFVR, pT=phosphothreonine**. Details are the same as for Supplementary Fig. 9.

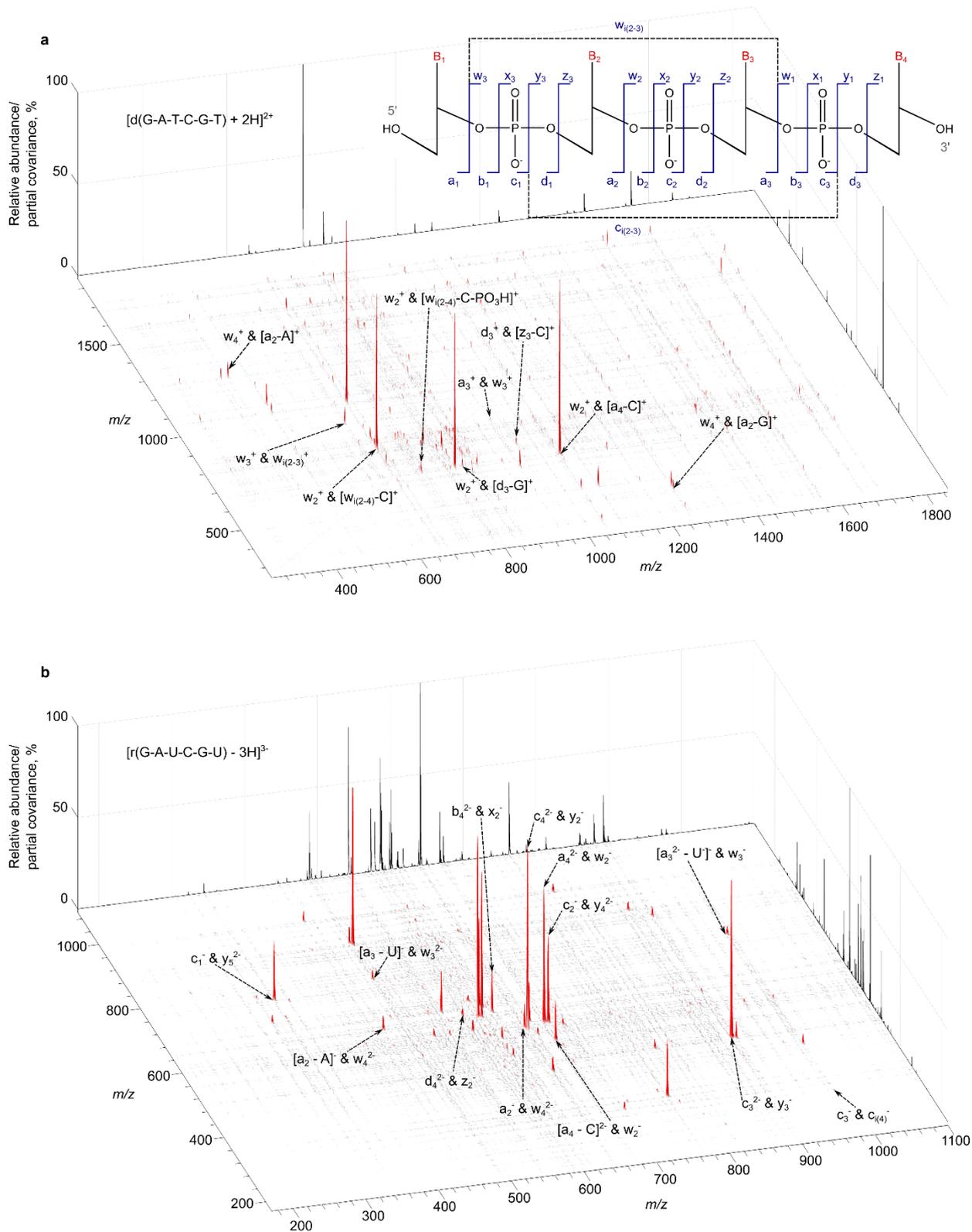

**Supplementary Figure 16 | pC-2DMS maps of DNA and RNA oligonucleotide ions. a,** doubly protonated DNA oligonucleotide [d(GATCGT)+2H]$^{2+}$ and **b,** triply charged negative ion of RNA oligonucleotide [r(GAUCGU)-3H]$^{3-}$. The inset is an illustration of the fragment ion nomenclature for oligonucleotide ions employed here. The terminal fragment nomenclature is according to Ref. 6*. Internal c-type fragments are annotated as $c_{i(n-m)}$ where n and m are the 5' and 3' nucleotides of the fragment sequence, respectively, or as $c_{i(n)}$ where the fragment is one nucleotide long, and the same subscript rules apply for internal w-type fragments. Rest of the details are as in Supplementary Fig. 9.

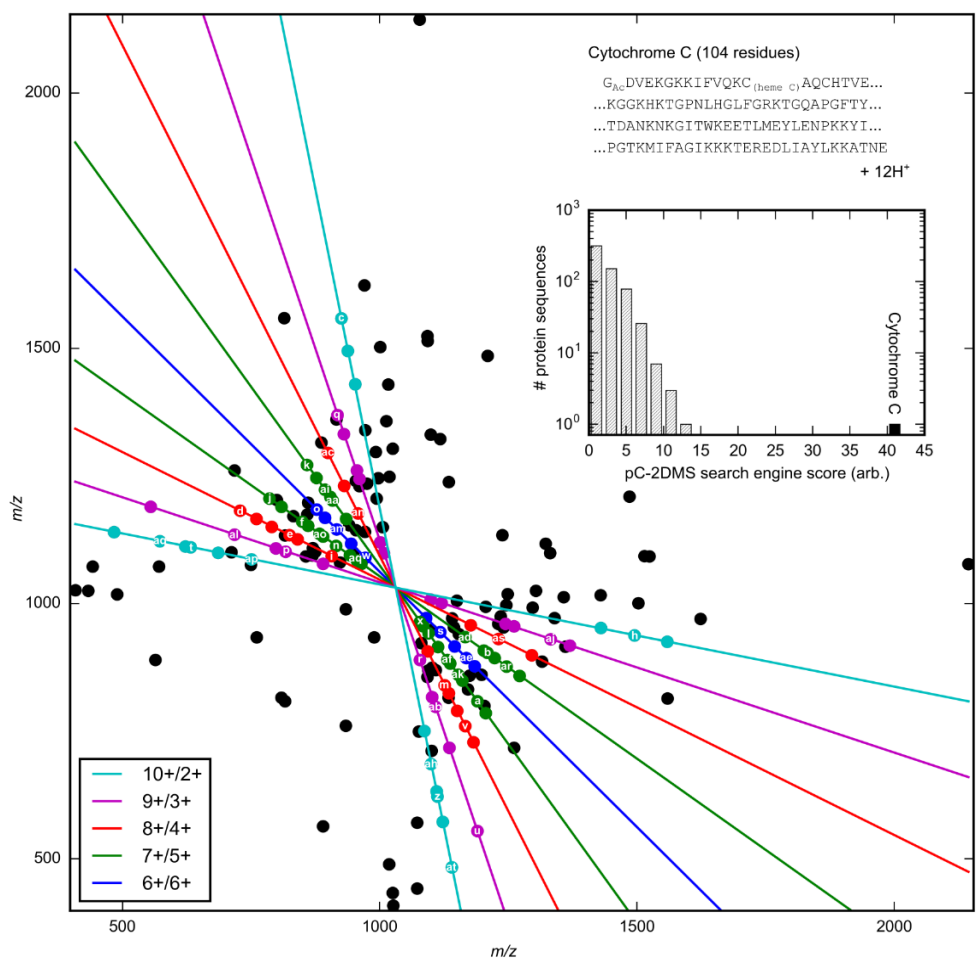

| Label | b/y complementary ion correlation | Label | b/y complementary ion correlation |
|---|---|---|---|
| a | $b_{69}^{7+}$ & $y_{35}^{5+}$ | x | $b_{63}^{7+}$ & $y_{41}^{5+}$ |
| b | $b_{50}^{5+}$ & $y_{54}^{7+}$ | y | $b_{75}^{9+}$ & $y_{29}^{3+}$ |
| c | $b_{22}^{2+}$ & $y_{82}^{10+}$ | z | $b_{11}^{2+}$ & $y_{93}^{10+}$ |
| d | $b_{79}^{8+}$ & $y_{25}^{4+}$ | aa | $b_{53}^{7+}$ & $y_{51}^{5+}$ |
| e | $b_{75}^{8+}$ & $y_{29}^{4+}$ | ab | $b_{84}^{9+}$ & $y_{20}^{3+}$ |
| f | $b_{33}^{5+}$ & $y_{71}^{7+}$ | ac | $b_{60}^{8+}$ & $y_{44}^{4+}$ |
| g | $b_{25}^{3+}$ & $y_{79}^{9+}$ | ad | $b_{48}^{5+}$ & $y_{56}^{7+}$ |
| h | $b_{21}^{2+}$ & $y_{83}^{10+}$ | ae | $b_{44}^{6+}$ & $y_{60}^{6+}$ |
| i | $b_{27}^{4+}$ & $y_{77}^{8+}$ | af | $b_{66}^{7+}$ & $y_{38}^{5+}$ |
| j | $b_{70}^{7+}$ & $y_{34}^{5+}$ | ag | $b_{38}^{5+}$ & $y_{66}^{7+}$ |
| k | $b_{53}^{5+}$ & $y_{51}^{7+}$ | ah | $b_{12}^{2+}$ & $y_{92}^{10+}$ |
| l | $b_{64}^{7+}$ & $y_{40}^{5+}$ | ai | $b_{51}^{5+}$ & $y_{53}^{7+}$ |
| m | $b_{25}^{4+}$ & $y_{79}^{8+}$ | aj | $b_{31}^{3+}$ & $y_{73}^{9+}$ |
| n | $b_{65}^{7+}$ & $y_{39}^{5+}$ | ak | $b_{34}^{5+}$ & $y_{70}^{7+}$ |
| o | $b_{43}^{6+}$ & $y_{61}^{6+}$ | al | $b_{86}^{9+}$ & $y_{18}^{3+}$ |
| p | $b_{83}^{9+}$ & $y_{21}^{3+}$ | am | $b_{58}^{6+}$ & $y_{46}^{6+}$ |
| q | $b_{32}^{3+}$ & $y_{72}^{9+}$ | an | $b_{64}^{8+}$ & $y_{40}^{4+}$ |
| r | $b_{81}^{9+}$ & $y_{23}^{3+}$ | ao | $b_{35}^{5+}$ & $y_{69}^{7+}$ |
| s | $b_{47}^{6+}$ & $y_{57}^{6+}$ | ap | $b_{13}^{2+}$ & $y_{91}^{10+}$ |
| t | $b_{93}^{10+}$ & $y_{11}^{2+}$ | aq | $b_{10}^{2+}$ & $y_{94}^{10+}$ |
| u | $b_{90}^{9+}$ & $y_{14}^{3+}$ | ar | $b_{52}^{5+}$ & $y_{52}^{7+}$ |
| v | $b_{78}^{8+}$ & $y_{26}^{4+}$ | as | $b_{62}^{8+}$ & $y_{42}^{4+}$ |
| w | $b_{48}^{6+}$ & $y_{56}^{6+}$ | at | $b_{96}^{10+}$ & $y_{8}^{2+}$ |

**Supplementary Figure 17 | pC-2DMS for top-down analysis of intact protein ions.** The scatter plot shows the top 100 correlation score-ranked pC-2DMS signals from CID on the 12+ ion of cytochrome c from equine heart (104 residues, ~12.4 kDa). The straight lines show the mass conservation lines for fragmentation of the intact molecule, with different colours denoting lines with different charge partitions between the complementary ions. The gradients and intercepts of the mass conservation lines provide the fragment charge states (see Eq. (S6)). The identities of the annotated b/y correlation signals are provided in the table, where the alphabetical order reflects the ranking according to pC-2DMS correlation score. By providing the measured pC-2DMS correlation signals to the pC-2DMS search engine, the correct sequence is automatically identified from database protein sequences by virtue of its outstanding pC-2DMS search engine score, as shown in the inset histogram.

| Fragment ion | P2 K5AcK16Ac | P3 K8AcK12Ac | P4 K5AcK12Ac | P5 K8AcK16Ac | Fragment ion | P2 K5AcK16Ac | P3 K8AcK12Ac | P4 K5AcK12Ac | P5 K8AcK16Ac |
|---|---|---|---|---|---|---|---|---|---|
| $[b/c/a]_2$ | x | o | x | o | $[b/c/a]_{i(4-5)}$ | x | o | x | o |
| $[b/c/a]_3$ | x | o | x | o | $[b/c/a]_{i(4-6)}$ | x | o | x | o |
| $[b/c/a]_4$ | x | o | x | o | $[b/c/a]_{i(4-7)}$ | x | o | x | o |
| $[b/c/a]_5$ | x | x | x | x | $[b/c/a]_{i(4-8)}$ | x | o | x | o |
| $[b/c/a]_6$ | x | x | x | x | $[b/c/a]_{i(4-9)}$ | ✓ | ✓ | x | x |
| $[b/c/a]_7$ | x | x | x | x | $[b/c/a]_{i(4-10)}$ | ✓ | ✓ | x | x |
| $[b/c/a]_8$ | x | x | x | x | $[b/c/a]_{i(4-11)}$ | ✓ | ✓ | x | x |
| $[b/c/a]_9$ | o | x | x | o | $[b/c/a]_{i(4-12)}$ | ✓ | ✓ | x | x |
| $[b/c/a]_{10}$ | o | x | x | o | $[b/c/a]_{i(4-13)}$ | x | o | x | o |
| $[b/c/a]_{11}$ | o | x | x | o | $[b/c/a]_{i(5-6)}$ | x | o | x | o |
| $[b/c/a]_{12}$ | o | x | x | o | $[b/c/a]_{i(5-7)}$ | x | o | x | o |
| $[b/c/a]_{13}$ | x | x | x | x | $[b/c/a]_{i(5-8)}$ | x | o | x | o |
| $[y/z/x]_1$ | x | x | x | x | $[b/c/a]_{i(5-9)}$ | ✓ | ✓ | x | x |
| $[y/z/x]_2$ | o | x | x | o | $[b/c/a]_{i(5-10)}$ | ✓ | ✓ | x | x |
| $[y/z/x]_3$ | o | x | x | o | $[b/c/a]_{i(5-11)}$ | ✓ | ✓ | x | x |
| $[y/z/x]_4$ | o | x | x | o | $[b/c/a]_{i(5-12)}$ | ✓ | ✓ | x | x |
| $[y/z/x]_5$ | o | x | x | o | $[b/c/a]_{i(5-13)}$ | x | o | x | o |
| $[y/z/x]_6$ | x | x | x | x | $[b/c/a]_{i(6-7)}$ | x | x | x | x |
| $[y/z/x]_7$ | x | x | x | x | $[b/c/a]_{i(6-8)}$ | x | x | x | x |
| $[y/z/x]_8$ | x | x | x | x | $[b/c/a]_{i(6-9)}$ | o | x | x | o |
| $[y/z/x]_9$ | x | x | x | x | $[b/c/a]_{i(6-10)}$ | o | x | x | o |
| $[y/z/x]_{10}$ | x | o | x | o | $[b/c/a]_{i(6-11)}$ | o | x | x | o |
| $[y/z/x]_{11}$ | x | o | x | o | $[b/c/a]_{i(6-12)}$ | o | x | x | o |
| $[y/z/x]_{12}$ | x | o | x | o | $[b/c/a]_{i(6-13)}$ | x | x | x | x |
| $[y/z/x]_{13}$ | x | x | x | x | $[b/c/a]_{i(7-8)}$ | x | x | x | x |
| $[b/c/a]_{i(2-3)}$ | x | o | x | o | $[b/c/a]_{i(7-9)}$ | o | x | x | o |
| $[b/c/a]_{i(2-4)}$ | x | o | x | o | $[b/c/a]_{i(7-10)}$ | o | x | x | o |
| $[b/c/a]_{i(2-5)}$ | x | x | x | x | $[b/c/a]_{i(7-11)}$ | o | x | x | o |
| $[b/c/a]_{i(2-6)}$ | x | x | x | x | $[b/c/a]_{i(7-12)}$ | o | x | x | o |
| $[b/c/a]_{i(2-7)}$ | x | x | x | x | $[b/c/a]_{i(7-13)}$ | x | x | x | x |
| $[b/c/a]_{i(2-8)}$ | x | x | x | x | $[b/c/a]_{i(8-9)}$ | o | x | x | o |
| $[b/c/a]_{i(2-9)}$ | o | x | x | o | $[b/c/a]_{i(8-10)}$ | o | x | x | o |
| $[b/c/a]_{i(2-10)}$ | o | x | x | o | $[b/c/a]_{i(8-11)}$ | o | x | x | o |
| $[b/c/a]_{i(2-11)}$ | o | x | x | o | $[b/c/a]_{i(8-12)}$ | o | x | x | o |
| $[b/c/a]_{i(2-12)}$ | o | x | x | o | $[b/c/a]_{i(8-13)}$ | x | x | x | x |
| $[b/c/a]_{i(2-13)}$ | x | x | x | x | $[b/c/a]_{i(9-10)}$ | o | x | x | o |
| $[b/c/a]_{i(3-4)}$ | x | x | x | x | $[b/c/a]_{i(9-11)}$ | o | x | x | o |
| $[b/c/a]_{i(3-5)}$ | x | o | x | o | $[b/c/a]_{i(9-12)}$ | o | x | x | o |
| $[b/c/a]_{i(3-6)}$ | x | o | x | o | $[b/c/a]_{i(9-13)}$ | x | x | x | x |
| $[b/c/a]_{i(3-7)}$ | x | o | x | o | $[b/c/a]_{i(10-11)}$ | x | x | x | x |
| $[b/c/a]_{i(3-8)}$ | x | o | x | o | $[b/c/a]_{i(10-12)}$ | x | x | x | x |
| $[b/c/a]_{i(3-9)}$ | ✓ | ✓ | x | x | $[b/c/a]_{i(10-13)}$ | o | x | x | o |
| $[b/c/a]_{i(3-10)}$ | ✓ | ✓ | x | x | $[b/c/a]_{i(11-12)}$ | x | x | x | x |
| $[b/c/a]_{i(3-11)}$ | ✓ | ✓ | x | x | $[b/c/a]_{i(11-13)}$ | o | x | x | o |
| $[b/c/a]_{i(3-12)}$ | ✓ | ✓ | x | x | $[b/c/a]_{i(12-13)}$ | o | x | x | o |
| $[b/c/a]_{i(3-13)}$ | x | o | x | o | | | | | |

**Supplementary Table 1 | All theoretically possible fragment ions that can be produced by all possible backbone cleavages of the four positional isomers P2–P5.** Rows denote all theoretically possible fragments (independent of fragmentation technique used, e.g. CID or ETD/ECD) that can be generated by one or multiple backbone cleavages of individual positional isomers (denoted by columns). The *m/z* values of the listed fragments are presented in Table 2. The symbols 'x' or 'o' across a row denote non-unique (identical or isomeric) fragments that can be produced from two or more positional isomers. The table demonstrates the absence of *any* unique 1D fragment, given any arbitrary mixture of all four isomers, for the positional isomers **P4** and **P5** under *any* theoretical backbone fragmentation. A tick denotes a unique 1D signal channel for **P2** and **P3**, however these amount to a selected few internal ions of which no 1D signal that could potentially correspond to such a fragment appears above ~2% RA in our standard 1D CID measurement of the P2-P5 mixture.

| Fragment ion | **P2** K5$_{Ac}$K16$_{Ac}$ | **P3** K8$_{Ac}$K12$_{Ac}$ | **P4** K5$_{Ac}$K12$_{Ac}$ | **P5** K8$_{Ac}$K16$_{Ac}$ |
|---|---|---|---|---|
| [b/c/a]$_2$ | 228.1343 / 243.1452 / 200.1394 | 186.1237 / 201.1346 / 158.1288 | 228.1343 / 243.1452 / 200.1394 | 186.1237 / 201.1346 / 158.1288 |
| [b/c/a]$_3$ | 285.1558 / 300.1666 / 257.1608 | 243.1452 / 258.156 / 215.1502 | 285.1558 / 300.1666 / 257.1608 | 243.1452 / 258.156 / 215.1502 |
| [b/c/a]$_4$ | 342.1772 / 357.1881 / 314.1823 | 300.1666 / 315.1775 / 272.1717 | 342.1772 / 357.1881 / 314.1823 | 300.1666 / 315.1775 / 272.1717 |
| [b/c/a]$_5$ | 470.2722 / 485.2831 / 442.2773 | 470.2722 / 485.2831 / 442.2773 | 470.2722 / 485.2831 / 442.2773 | 470.2722 / 485.2831 / 442.2773 |
| [b/c/a]$_6$ | 527.2936 / 542.3045 / 499.2987 | 527.2936 / 542.3045 / 499.2987 | 527.2936 / 542.3045 / 499.2987 | 527.2936 / 542.3045 / 499.2987 |
| [b/c/a]$_7$ | 640.3777 / 655.3886 / 612.3828 | 640.3777 / 655.3886 / 612.3828 | 640.3777 / 655.3886 / 612.3828 | 640.3777 / 655.3886 / 612.3828 |
| [b/c/a]$_8$ | 697.3992 / 712.4101 / 669.4042 | 697.3992 / 712.4101 / 669.4042 | 697.3992 / 712.4101 / 669.4042 | 697.3992 / 712.4101 / 669.4042 |
| [b/c/a]$_9$ | 825.4941 / 840.505 / 797.4992 | 867.5047 / 882.5156 / 839.5098 | 867.5047 / 882.5156 / 839.5098 | 825.4941 / 840.505 / 797.4992 |
| [b/c/a]$_{10}$ | 882.5156 / 897.5265 / 854.5207 | 924.5262 / 939.5371 / 896.5313 | 924.5262 / 939.5371 / 896.5313 | 882.5156 / 897.5265 / 854.5207 |
| [b/c/a]$_{11}$ | 939.537 / 954.5479 / 911.5421 | 981.5476 / 996.5585 / 953.5527 | 981.5476 / 996.5585 / 953.5527 | 939.537 / 954.5479 / 911.5421 |
| [b/c/a]$_{12}$ | 1010.5742 / 1025.585 / 982.5792 | 1052.5848 / 1067.5956 / 1024.5898 | 1052.5848 / 1067.5956 / 1024.5898 | 1010.5742 / 1025.585 / 982.5792 |
| [b/c/a]$_{13}$ | 1180.6797 / 1195.6906 / 1152.6848 | 1180.6797 / 1195.6906 / 1152.6848 | 1180.6797 / 1195.6906 / 1152.6848 | 1180.6797 / 1195.6906 / 1152.6848 |
| [y/z/x]$_1$ | 175.119 / 160.1081 / 203.1139 | 175.119 / 160.1081 / 203.1139 | 175.119 / 160.1081 / 203.1139 | 175.119 / 160.1081 / 203.1139 |
| [y/z/x]$_2$ | 345.2245 / 330.2136 / 373.2194 | 303.2139 / 288.203 / 331.2088 | 303.2139 / 288.203 / 331.2088 | 345.2245 / 330.2136 / 373.2194 |
| [y/z/x]$_3$ | 416.2616 / 401.2507 / 444.2565 | 374.251 / 359.2401 / 402.2459 | 374.251 / 359.2401 / 402.2459 | 416.2616 / 401.2507 / 444.2565 |
| [y/z/x]$_4$ | 473.2831 / 458.2722 / 501.278 | 431.2725 / 416.2616 / 459.2674 | 431.2725 / 416.2616 / 459.2674 | 473.2831 / 458.2722 / 501.278 |
| [y/z/x]$_5$ | 530.3045 / 515.2936 / 558.2995 | 488.2939 / 473.283 / 516.2889 | 488.2939 / 473.283 / 516.2889 | 530.3045 / 515.2936 / 558.2995 |
| [y/z/x]$_6$ | 658.3995 / 643.3886 / 686.3944 | 658.3995 / 643.3886 / 686.3944 | 658.3995 / 643.3886 / 686.3944 | 658.3995 / 643.3886 / 686.3944 |
| [y/z/x]$_7$ | 715.421 / 700.4101 / 743.4159 | 715.421 / 700.4101 / 743.4159 | 715.421 / 700.4101 / 743.4159 | 715.421 / 700.4101 / 743.4159 |
| [y/z/x]$_8$ | 828.505 / 813.4941 / 856.4999 | 828.505 / 813.4941 / 856.4999 | 828.505 / 813.4941 / 856.4999 | 828.505 / 813.4941 / 856.4999 |
| [y/z/x]$_9$ | 885.5265 / 870.5156 / 913.5214 | 885.5265 / 870.5156 / 913.5214 | 885.5265 / 870.5156 / 913.5214 | 885.5265 / 870.5156 / 913.5214 |
| [y/z/x]$_{10}$ | 1013.6214 / 998.6106 / 1041.6164 | 1055.632 / 1040.6212 / 1083.627 | 1013.6214 / 998.6106 / 1041.6164 | 1055.632 / 1040.6212 / 1083.627 |
| [y/z/x]$_{11}$ | 1070.6429 / 1055.632 / 1098.6378 | 1112.6535 / 1097.6426 / 1140.6484 | 1070.6429 / 1055.632 / 1098.6378 | 1112.6535 / 1097.6426 / 1140.6484 |
| [y/z/x]$_{12}$ | 1127.6644 / 1112.6535 / 1155.6593 | 1169.675 / 1154.6641 / 1197.6699 | 1127.6644 / 1112.6535 / 1155.6593 | 1169.675 / 1154.6641 / 1197.6699 |
| [y/z/x]$_{13}$ | 1297.7699 / 1282.759 / 1325.7648 | 1297.7699 / 1282.759 / 1325.7648 | 1297.7699 / 1282.759 / 1325.7648 | 1297.7699 / 1282.759 / 1325.7648 |
| [b/c/a]$_{i(2-3)}$ | 228.1343 / 243.1452 / 200.1394 | 186.1237 / 201.1346 / 158.1288 | 228.1343 / 243.1452 / 200.1394 | 186.1237 / 201.1346 / 158.1288 |
| [b/c/a]$_{i(2-4)}$ | 285.1558 / 300.1666 / 257.1608 | 243.1452 / 258.156 / 215.1502 | 285.1558 / 300.1666 / 257.1608 | 243.1452 / 258.156 / 215.1502 |
| [b/c/a]$_{i(2-5)}$ | 413.2507 / 428.2616 / 385.2558 | 413.2507 / 428.2616 / 385.2558 | 413.2507 / 428.2616 / 385.2558 | 413.2507 / 428.2616 / 385.2558 |
| [b/c/a]$_{i(2-6)}$ | 470.2722 / 485.2831 / 442.2773 | 470.2722 / 485.2831 / 442.2773 | 470.2722 / 485.2831 / 442.2773 | 470.2722 / 485.2831 / 442.2773 |
| [b/c/a]$_{i(2-7)}$ | 583.3562 / 598.3671 / 555.3613 | 583.3562 / 598.3671 / 555.3613 | 583.3562 / 598.3671 / 555.3613 | 583.3562 / 598.3671 / 555.3613 |
| [b/c/a]$_{i(2-8)}$ | 640.3777 / 655.3886 / 612.3828 | 640.3777 / 655.3886 / 612.3828 | 640.3777 / 655.3886 / 612.3828 | 640.3777 / 655.3886 / 612.3828 |
| [b/c/a]$_{i(2-9)}$ | 768.4727 / 783.4836 / 740.4777 | 810.4833 / 825.4942 / 782.4883 | 810.4833 / 825.4942 / 782.4883 | 768.4727 / 783.4836 / 740.4777 |
| [b/c/a]$_{i(2-10)}$ | 825.4941 / 840.505 / 797.4992 | 867.5047 / 882.5156 / 839.5098 | 867.5047 / 882.5156 / 839.5098 | 825.4941 / 840.505 / 797.4992 |
| [b/c/a]$_{i(2-11)}$ | 882.5156 / 897.5265 / 854.5207 | 924.5262 / 939.5371 / 896.5313 | 924.5262 / 939.5371 / 896.5313 | 882.5156 / 897.5265 / 854.5207 |
| [b/c/a]$_{i(2-12)}$ | 953.5527 / 968.5636 / 925.5578 | 995.5633 / 1010.5742 / 967.5684 | 995.5633 / 1010.5742 / 967.5684 | 953.5527 / 968.5636 / 925.5578 |
| [b/c/a]$_{i(2-13)}$ | 1123.6583 / 1138.6691 / 1095.6633 | 1123.6582 / 1138.6691 / 1095.6633 | 1123.6582 / 1138.6691 / 1095.6633 | 1123.6583 / 1138.6691 / 1095.6633 |
| [b/c/a]$_{i(3-4)}$ | 115.0502 / 130.0611 / 87.0553 | 115.0502 / 130.0611 / 87.0553 | 115.0502 / 130.0611 / 87.0553 | 115.0502 / 130.0611 / 87.0553 |
| [b/c/a]$_{i(3-5)}$ | 243.1452 / 258.156 / 215.1502 | 285.1558 / 300.1666 / 257.1608 | 243.1452 / 258.156 / 215.1502 | 285.1558 / 300.1666 / 257.1608 |
| [b/c/a]$_{i(3-6)}$ | 300.1666 / 315.1775 / 272.1717 | 342.1772 / 357.1881 / 314.1823 | 300.1666 / 315.1775 / 272.1717 | 342.1772 / 357.1881 / 314.1823 |
| [b/c/a]$_{i(3-7)}$ | 413.2507 / 428.2616 / 385.2558 | 455.2613 / 470.2722 / 427.2664 | 413.2507 / 428.2616 / 385.2558 | 455.2613 / 470.2722 / 427.2664 |
| [b/c/a]$_{i(3-8)}$ | 470.2721 / 485.283 / 442.2772 | 512.2827 / 527.2936 / 484.2878 | 470.2721 / 485.283 / 442.2772 | 512.2827 / 527.2936 / 484.2878 |
| [b/c/a]$_{i(3-9)}$ | 598.3671 / 613.378 / 570.3722 | 682.3883 / 697.3992 / 654.3934 | 640.3777 / 655.3886 / 612.3828 | 640.3777 / 655.3886 / 612.3828 |
| [b/c/a]$_{i(3-10)}$ | 655.3886 / 670.3994 / 627.3936 | 739.4098 / 754.4206 / 711.4148 | 697.3992 / 712.4101 / 669.4042 | 697.3992 / 712.41 / 669.4042 |
| [b/c/a]$_{i(3-11)}$ | 712.41 / 727.4209 / 684.4151 | 796.4312 / 811.4421 / 768.4363 | 754.4206 / 769.4315 / 726.4257 | 754.4206 / 769.4315 / 726.4257 |
| [b/c/a]$_{i(3-12)}$ | 783.4471 / 798.458 / 755.4522 | 867.4683 / 882.4792 / 839.4734 | 825.4577 / 840.4686 / 797.4628 | 825.4577 / 840.4686 / 797.4628 |
| [b/c/a]$_{i(3-13)}$ | 953.5527 / 968.5636 / 925.5578 | 995.5633 / 1010.5742 / 967.5684 | 953.5527 / 968.5636 / 925.5578 | 995.5633 / 1010.5742 / 967.5684 |
| [b/c/a]$_{i(4-5)}$ | 186.1237 / 201.1346 / 158.1288 | 228.1343 / 243.1452 / 200.1394 | 186.1237 / 201.1346 / 158.1288 | 228.1343 / 243.1452 / 200.1394 |
| [b/c/a]$_{i(4-6)}$ | 243.1452 / 258.156 / 215.1502 | 285.1558 / 300.1666 / 257.1608 | 243.1452 / 258.156 / 215.1502 | 285.1558 / 300.1666 / 257.1608 |
| [b/c/a]$_{i(4-7)}$ | 356.2292 / 371.2401 / 328.2343 | 398.2398 / 413.2507 / 370.2449 | 356.2292 / 371.2401 / 328.2343 | 398.2398 / 413.2507 / 370.2449 |
| [b/c/a]$_{i(4-8)}$ | 413.2507 / 428.2616 / 385.2558 | 455.2613 / 470.2722 / 427.2664 | 413.2507 / 428.2616 / 385.2558 | 455.2613 / 470.2722 / 427.2664 |
| [b/c/a]$_{i(4-9)}$ | 541.3456 / 556.3565 / 513.3507 | 625.3668 / 640.3777 / 597.3719 | 583.3562 / 598.3671 / 555.3613 | 583.3562 / 598.3671 / 555.3613 |
| [b/c/a]$_{i(4-10)}$ | 598.3671 / 613.378 / 570.3722 | 682.3883 / 697.3992 / 654.3934 | 640.3777 / 655.3886 / 612.3828 | 640.3777 / 655.3886 / 612.3828 |
| [b/c/a]$_{i(4-11)}$ | 655.3886 / 670.3995 / 627.3936 | 739.4098 / 754.4206 / 711.4148 | 697.3992 / 712.4101 / 669.4042 | 697.3992 / 712.4101 / 669.4042 |
| [b/c/a]$_{i(4-12)}$ | 726.4257 / 741.4366 / 698.4308 | 810.4469 / 825.4578 / 782.452 | 768.4363 / 783.4472 / 740.4414 | 768.4363 / 783.4472 / 740.4414 |
| [b/c/a]$_{i(4-13)}$ | 896.5312 / 911.5421 / 868.5363 | 938.5418 / 953.5527 / 910.5469 | 896.5312 / 911.5421 / 868.5363 | 938.5418 / 953.5527 / 910.5469 |
| [b/c/a]$_{i(5-6)}$ | 186.1237 / 201.1346 / 158.1288 | 228.1343 / 243.1452 / 200.1394 | 186.1237 / 201.1346 / 158.1288 | 228.1343 / 243.1452 / 200.1394 |
| [b/c/a]$_{i(5-7)}$ | 299.2078 / 314.2186 / 271.2128 | 341.2184 / 356.2292 / 313.2234 | 299.2078 / 314.2186 / 271.2128 | 341.2184 / 356.2292 / 313.2234 |
| [b/c/a]$_{i(5-8)}$ | 356.2292 / 371.2401 / 328.2343 | 398.2398 / 413.2507 / 370.2449 | 356.2292 / 371.2401 / 328.2343 | 398.2398 / 413.2507 / 370.2449 |
| [b/c/a]$_{i(5-9)}$ | 484.3242 / 499.3351 / 456.3293 | 568.3454 / 583.3563 / 540.3505 | 526.3348 / 541.3457 / 498.3399 | 526.3348 / 541.3457 / 498.3399 |
| [b/c/a]$_{i(5-10)}$ | 541.3456 / 556.3565 / 513.3507 | 625.3668 / 640.3777 / 597.3719 | 583.3562 / 598.3671 / 555.3613 | 583.3562 / 598.3671 / 555.3613 |

**Supplementary Table 2 | Mass-to-charge ratios of all theoretically possible fragments that can be produced by all possible backbone cleavages of the four positional isomers P2–P5.** Rows denote the *m/z* values of all theoretically possible fragments that can be generated (independently of fragmentation technique used, e.g. CID or ETD/ECD) following one or multiple backbone cleavages of the individual positional isomers (denoted by columns).

| Fragment ion | **P2**<br>K5$_{Ac}$K16$_{Ac}$ | **P3**<br>K8$_{Ac}$K12$_{Ac}$ | **P4**<br>K5$_{Ac}$K12$_{Ac}$ | **P5**<br>K8$_{Ac}$K16$_{Ac}$ |
|---|---|---|---|---|
| [b/c/a]$_{i(5-11)}$ | 598.3671 / 613.378 / 570.3722 | 682.3883 / 697.3992 / 654.3934 | 640.3777 / 655.3886 / 612.3828 | 640.3777 / 655.3886 / 612.3828 |
| [b/c/a]$_{i(5-12)}$ | 669.4042 / 684.4151 / 641.4093 | 753.4254 / 768.4363 / 725.4305 | 711.4148 / 726.4257 / 683.4199 | 711.4148 / 726.4257 / 683.4199 |
| [b/c/a]$_{i(5-13)}$ | 839.5098 / 854.5207 / 811.5148 | 881.5204 / 896.5313 / 853.5254 | 839.5098 / 854.5207 / 811.5148 | 881.5204 / 896.5313 / 853.5254 |
| [b/c/a]$_{i(6-7)}$ | 171.1128 / 186.1237 / 143.1179 | 171.1128 / 186.1237 / 143.1179 | 171.1128 / 186.1237 / 143.1179 | 171.1128 / 186.1237 / 143.1179 |
| [b/c/a]$_{i(6-8)}$ | 228.1343 / 243.1452 / 200.1393 | 228.1343 / 243.1452 / 200.1393 | 228.1343 / 243.1452 / 200.1393 | 228.1343 / 243.1452 / 200.1393 |
| [b/c/a]$_{i(6-9)}$ | 356.2292 / 371.2401 / 328.2343 | 398.2398 / 413.2507 / 370.2449 | 398.2398 / 413.2507 / 370.2449 | 356.2292 / 371.2401 / 328.2343 |
| [b/c/a]$_{i(6-10)}$ | 413.2507 / 428.2616 / 385.2558 | 455.2613 / 470.2722 / 427.2664 | 455.2613 / 470.2722 / 427.2664 | 413.2507 / 428.2616 / 385.2558 |
| [b/c/a]$_{i(6-11)}$ | 470.2721 / 485.283 / 442.2772 | 512.2827 / 527.2936 / 484.2878 | 512.2827 / 527.2936 / 484.2878 | 470.2721 / 485.283 / 442.2772 |
| [b/c/a]$_{i(6-12)}$ | 541.3092 / 556.3201 / 513.3143 | 583.3198 / 598.3307 / 555.3249 | 583.3198 / 598.3307 / 555.3249 | 541.3092 / 556.3201 / 513.3143 |
| [b/c/a]$_{i(6-13)}$ | 711.4148 / 726.4257 / 683.4199 | 711.4148 / 726.4257 / 683.4199 | 711.4148 / 726.4257 / 683.4199 | 711.4148 / 726.4257 / 683.4199 |
| [b/c/a]$_{i(7-8)}$ | 171.1128 / 186.1237 / 143.1179 | 171.1128 / 186.1237 / 143.1179 | 171.1128 / 186.1237 / 143.1179 | 171.1128 / 186.1237 / 143.1179 |
| [b/c/a]$_{i(7-9)}$ | 299.2078 / 314.2186 / 271.2128 | 341.2184 / 356.2292 / 313.2234 | 341.2184 / 356.2292 / 313.2234 | 299.2078 / 314.2186 / 271.2128 |
| [b/c/a]$_{i(7-10)}$ | 356.2292 / 371.2401 / 328.2343 | 398.2398 / 413.2507 / 370.2449 | 398.2398 / 413.2507 / 370.2449 | 356.2292 / 371.2401 / 328.2343 |
| [b/c/a]$_{i(7-11)}$ | 413.2507 / 428.2616 / 385.2558 | 455.2613 / 470.2722 / 427.2664 | 455.2613 / 470.2722 / 427.2664 | 413.2507 / 428.2616 / 385.2558 |
| [b/c/a]$_{i(7-12)}$ | 484.2878 / 499.2987 / 456.2929 | 526.2984 / 541.3093 / 498.3035 | 526.2984 / 541.3093 / 498.3035 | 484.2878 / 499.2987 / 456.2929 |
| [b/c/a]$_{i(7-13)}$ | 654.3933 / 669.4042 / 626.3984 | 654.3933 / 669.4042 / 626.3984 | 654.3933 / 669.4042 / 626.3984 | 654.3933 / 669.4042 / 626.3984 |
| [b/c/a]$_{i(8-9)}$ | 186.1237 / 201.1346 / 158.1288 | 228.1343 / 243.1452 / 200.1394 | 228.1343 / 243.1452 / 200.1394 | 186.1237 / 201.1346 / 158.1288 |
| [b/c/a]$_{i(8-10)}$ | 243.1452 / 258.156 / 215.1502 | 285.1558 / 300.1666 / 257.1608 | 285.1558 / 300.1666 / 257.1608 | 243.1452 / 258.156 / 215.1502 |
| [b/c/a]$_{i(8-11)}$ | 300.1666 / 315.1775 / 272.1717 | 342.1772 / 357.1881 / 314.1823 | 342.1772 / 357.1881 / 314.1823 | 300.1666 / 315.1775 / 272.1717 |
| [b/c/a]$_{i(8-12)}$ | 371.2037 / 386.2146 / 343.2088 | 413.2143 / 428.2252 / 385.2194 | 413.2143 / 428.2252 / 385.2194 | 371.2037 / 386.2146 / 343.2088 |
| [b/c/a]$_{i(8-13)}$ | 541.3093 / 556.3202 / 513.3144 | 541.3093 / 556.3202 / 513.3144 | 541.3093 / 556.3202 / 513.3144 | 541.3093 / 556.3202 / 513.3144 |
| [b/c/a]$_{i(9-10)}$ | 186.1237 / 201.1346 / 158.1288 | 228.1343 / 243.1452 / 200.1394 | 228.1343 / 243.1452 / 200.1394 | 186.1237 / 201.1346 / 158.1288 |
| [b/c/a]$_{i(9-11)}$ | 243.1452 / 258.156 / 215.1502 | 285.1558 / 300.1666 / 257.1608 | 285.1558 / 300.1666 / 257.1608 | 243.1452 / 258.156 / 215.1502 |
| [b/c/a]$_{i(9-12)}$ | 314.1823 / 329.1932 / 286.1874 | 356.1929 / 371.2038 / 328.1979 | 356.1929 / 371.2038 / 328.1979 | 314.1823 / 329.1932 / 286.1874 |
| [b/c/a]$_{i(9-13)}$ | 484.2878 / 499.2987 / 456.2929 | 484.2878 / 499.2987 / 456.2929 | 484.2878 / 499.2987 / 456.2929 | 484.2878 / 499.2987 / 456.2929 |
| [b/c/a]$_{i(10-11)}$ | 115.0502 / 130.0611 / 87.0553 | 115.0502 / 130.0611 / 87.0553 | 115.0502 / 130.0611 / 87.0553 | 115.0502 / 130.0611 / 87.0553 |
| [b/c/a]$_{i(10-12)}$ | 186.0873 / 201.0982 / 158.0924 | 186.0873 / 201.0982 / 158.0924 | 186.0873 / 201.0982 / 158.0924 | 186.0873 / 201.0982 / 158.0924 |
| [b/c/a]$_{i(10-13)}$ | 356.1929 / 371.2038 / 328.1979 | 314.1823 / 329.1932 / 286.1874 | 314.1823 / 329.1932 / 286.1874 | 356.1929 / 371.2038 / 328.1979 |
| [b/c/a]$_{i(11-12)}$ | 129.0659 / 144.0767 / 101.0709 | 129.0659 / 144.0767 / 101.0709 | 129.0659 / 144.0767 / 101.0709 | 129.0659 / 144.0767 / 101.0709 |
| [b/c/a]$_{i(11-13)}$ | 299.1714 / 314.1823 / 271.1765 | 257.1608 / 272.1717 / 229.1659 | 257.1608 / 272.1717 / 229.1659 | 299.1714 / 314.1823 / 271.1765 |
| [b/c/a]$_{i(12-13)}$ | 242.1499 / 257.1608 / 214.155 | 200.1394 / 215.1502 / 172.1444 | 200.1394 / 215.1502 / 172.1444 | 242.1499 / 257.1608 / 214.155 |

**Supplementary Table 2 (continued) | Mass-to-charge ratios of all theoretically possible fragments that can be produced by all possible backbone cleavages of the four positional isomers P2–P5**. Rows denote the *m/z* values of all theoretically possible fragments that can be generated (independently of fragmentation technique used, e.g. CID or ETD/ECD) following one or multiple backbone cleavages of the individual positional isomers (denoted by columns).

| Sequence | Modification | Charge states analysed |
|---|---|---|
| **SApTPEALAFVR** | pT=phosphothreonine | 2+ |
| **TWpTLCGTVEY** | pT=phosphothreonine | 2+ |
| **RDpYTGW-Nle-DF-NH$_2$** | pY=phosphotyrosine | 2+, 2− |
| **DsYMGWMDF-NH$_2$** | sY=sulphotyrosine | 2+ |
| **EQFDDsYGHMRF-NH$_2$** | sY=sulphotyrosine | 2+ |
| **RDsYTGW-Nle-DF-NH$_2$** | sY=sulphotyrosine | 2+ |
| **GDFEEIPEEYLQ** | None | 2− |
| **VTIMPK(Ac)DIQLAR** | K(Ac)= $N^\varepsilon$-*acetyllysine* | 2+, 3+ |
| **GGNFSGR(Me)GGFGGSR** | R(Me)=$N^G$-monomethylarginine | 2+, 3+ |
| **GWGR(Me$_2$)EENLFSWK** | R(Me$_2$)=$N^G$,$N^{G'}$-dimethylarginine | 2+, 3+, 2− |
| **TWR(Me$_2$)GGEEK** | R(Me$_2$)= $N^G$,$N^G$-dimethylarginine | 2+, 3+ |
| **VTIMPK(Me$_3$)DIQLAR** | K(Me$_3$)= $N^\varepsilon$,$N^\varepsilon$,$N^\varepsilon$-*trimethyllysine* | 2+, 3+ |
| **LGEY(nitro)GFQNAILVR** | Y(nitro)=3-nitrotyrosine | 2+, 3+ |
| **EQFDDpYGHMRF-NH$_2$** | pY=phosphotyrosine | 2+, 3+ |
| **GSNKGAIIGLM** | None | 2+, 3+ |
| **MLGIIAGKNSG** | None | 2+, 3+ |
| **VTIMPKDIQLAR** | None | 2+, 3+ |
| **YGGFLRRIRPKLK** | None | 2+, 3+ |
| **GK(Ac)GGKGLGK(Ac)GGAKR** | K(Ac)= $N^\varepsilon$-*acetyllysine* | 2+, 3+ |
| **GKGGK(Ac)GLGKGGAK(Ac)R** | K(Ac)= $N^\varepsilon$-*acetyllysine* | 2+, 3+ |
| **GK(Ac)GGKGLGKGGAK(Ac)R** | K(Ac)= $N^\varepsilon$-*acetyllysine* | 2+, 3+ |
| **GKGGK(Ac)GLGK(Ac)GGAKR** | K(Ac)= $N^\varepsilon$-*acetyllysine* | 2+, 3+ |
| **GDFEEIPEEpYLQ** | pY=phosphotyrosine | 2−, 3− |
| **d(5'–GATCGT–3')** | None | 2+ |
| **r(5'–GAUCGU–3')** | None | 2−, 3−, 4− |

**Supplementary Table 3 | List of modified and unmodified peptides as well as DNA and RNA oligonucleotide sequences used for validation of pC-2DMS**. The tested parent ion charge states are provided for each analysed sequence. For each examined parent ion, the pC-2DMS map was generated, the scored list of correlating fragments was produced and the high-scoring correlations were confirmed to be free of any extrinsic (structure-unrelated) correlations.

# References[a]:

[a]Citations of the Supplementary Information references are marked with *asterisk* (*) throughout.